\documentclass[multphys,vecphys]{svmult}

\usepackage{amssymb}
\usepackage[usenames]{color}
\usepackage{amsmath}
\usepackage{makeidx}         
\usepackage{graphicx}        
\usepackage{multicol}        
\usepackage{cite}            
\usepackage[bottom]{footmisc}

\makeindex             

\newcommand {\ET}{ET }

\newcommand {\dmit}{$\beta'$-[Pd\-(dmit)$_2$]$_2\-Z$ }
\newcommand {\dmitn}{$\beta'$-[Pd\-(dmit)$_2$]$_2\-Z$}

\newcommand {\Br}{$\kappa$-(ET)$_2$\-Cu\-[N\-(CN)$_2$]\-Br }
\newcommand {\Brn}{$\kappa$-(ET)$_2$\-Cu\-[N\-(CN)$_2$]\-Br}
\newcommand {\Cl}{$\kappa$-(ET)$_2$\-Cu\-[N\-(CN)$_2$]\-Cl }
\newcommand {\NCS}{$\kappa$-(ET)$_2$\-Cu\-(NCS)$_2$ }
\newcommand {\Ga}{$\lambda$-(BETS)$_2$\-Ga\-Cl$_4$ }
\newcommand {\kFe}{$\kappa$-(BETS)$_2$\-Fe\-Br$_4$ }

\newcommand {\bI}{$\beta$-(ET)$_2$\-I$_3$ }

\newcommand {\CN}{$\kappa$-(ET)$_2$\-Cu$_2$\-(CN)$_3$ }
\newcommand {\CNn}{$\kappa$-(ET)$_2\-$Cu$_2$\-(CN)$_3$}
\newcommand {\Cln}{$\kappa$-(ET)$_2$\-Cu\-[N\-(CN)$_2$]\-Cl}
\newcommand {\NCSn}{$\kappa$-(ET)$_2$\-Cu\-(NCS)$_2$}

\newcommand {\SRO}{Sr$_2$\-Ru\-O$_4$ }
\newcommand {\SROn}{Sr$_2$\-Ru\-O$_4$}
\newcommand {\YBCO}{YBa$_2$\-Cu$_3$O$_{7-x}$ }
\newcommand {\YBCOn}{YBa$_2$\-Cu$_3$O$_{7-x}$}

\newcommand {\etal}{{\it et al}. }
\newcommand {\etaln}{{\it et al}.}

\begin{document}

\title{Strong electronic correlations in superconducting organic charge transfer salts.}
\titlerunning{Strong electronic correlations in organic charge
transfer salts}
\author{B. J. Powell \and Ross H. McKenzie}
\institute{Department of Physics, University of Queensland,
Brisbane, Queensland 4072, Australia.
\texttt{powell@physics.uq.edu.au}}


\maketitle

\begin{abstract}

We review the role of strong electronic correlations in
quasi--two-dimensional organic charge transfer salts such as
(BEDT-TTF)$_2X$, (BETS)$_2Y$ and $\beta'$-[Pd(dmit)$_2$]$_2Z$. We
begin by defining minimal models for these materials. It is
necessary to identify two classes of material: the first class is
strongly dimerised and is described by a half-filled Hubbard model;
the second class is not strongly dimerised and is described by a
quarter filled extended Hubbard model. We argue that these models
capture the essential physics of these materials. We explore the
phase diagram of the half-filled quasi--two-dimensional organic
charge transfer salts, focusing on the metallic and superconducting
phases. We review work showing that the metallic phase, which has
both Fermi liquid and `bad metal' regimes, is described both
quantitatively and qualitatively by dynamical mean field theory
(DMFT). The phenomenology of the superconducting state is still a
matter of contention. We critically review the experimental
situation, focusing on the key experimental results that may
distinguish between rival theories of superconductivity,
particularly probes of the pairing symmetry and measurements of the
superfluid stiffness. We then discuss some strongly correlated
theories of superconductivity, in particular, the resonating valence
bond (RVB) theory of superconductivity. We conclude by discussing
some of the major challenges currently facing the field. These
include: parameterising minimal models; the evidence for a pseudogap
from nuclear magnetic resonance (NMR) experiments; superconductors
with low critical temperatures and extremely small superfluid
stiffnesses; the possible spin-liquid states in \CN and \dmitn; and
the need for high quality large single crystals.
\end{abstract}

%
%
%
%
%
%
%

\section{Motivation and scope}

The Drude, Sommerfeld and Bloch models explain many of the
properties of elemental metals and alloys in terms of a theory of
non-interacting electrons. This is somewhat surprising as for
typical metals the radius of a sphere whose volume is equal to the
volume per conduction electron is of order 1~\AA \cite{A&V}. Thus,
one na\"ively expects that the Coulomb interaction between electrons
will be large. Therefore, one of the most significant achievements
in condensed matter theory is Landau's theory of Fermi liquids
\cite{Landau} which explains, through the idea of adiabatic
continuity, why the Coulomb interation simply renormalises the
weakly interacting electron gas.

However, in many ``strongly correlated'' systems interactions
qualitatively alter the behaviour of the material. For example, in
the Kondo effect the interactions between the conduction electrons
and magnetic impurities, which leads to a logarithmic increase in
the resistivity at low temperatures, cannot be described by
perturbation theory from the Fermi liquid state \cite{Hewson}.
Indeed even the BCS state \cite{BCS} is not adiabatically connected
to the Fermi liquid ground state. The Kondo effect is now well
understood, but many other strongly correlated phases are not, e.g.,
high temperature superconductivity and the pseudogap in the cuprates
\cite{LeeWenNagaosa}, colossal magnetoresistance in the manganites
\cite{Salamon} and emergence of unconventional superconductivity in
heavy fermion materials \cite{Sigrist&Ueda}.

The non-perturbative nature of strongly correlated materials
presents several major challenges to theory. For example some big
questions are: What is the appropriate description of the metallic
state when the system is near to an instability to a Mott insulating
state? What are the connections between magnetic ordering and
unconventional superconductivity? Under what conditions can an
antiferromagnet have a spin-liquid ground state and/or spinon
excitations? When superconductivity is found near a Mott transition,
what is the relationship between the symmetry of the superconducting
state and the ground state of the parent Mott insulator? What is the
appropriate microscopic description of a metallic state with a
pseudogap? What are the mechanisms responsible for unconventional
superconductivity and what are the appropriate microscopic
descriptions of such states? Many of these questions are deeply
intertwined.

Organic charge transfer salts are excellent model systems in which
to study many of the questions about the strongly correlated
phenomena described above. Band structure suggests that these
materials should be metals at all experimentally relevant pressures
and temperatures. However, as we will discuss below, the observed
phases include, Mott insulators, N\'eel antiferromagnets,
spin-liquids, (unconventional) superconductors, Fermi liquids, a
pseudogap and a `bad metal'. Further, organic chemistry allows the
organic charge transfer salts to be subtly tuned in ways that have
never been achieved in inorganic materials. A rather beautiful
example of this is that in \Br the Mott transition can be driven by
replacing the eight hydrogen atoms in the \ET molecule with
deuterium \cite{Taniguchi-Mott}. Indeed the chemistry can be
controlled to such an extent that the number of H/D atoms on each
\ET molecule can be varied uniformly throughout the entire sample
allowing one to move gradually across the Mott transition and
observe the coexistence of the insulating and superconducting phases
expected because the transition is first order \cite{Sasaki}.

We will argue below that, despite the chemical complexity of the
organic charge transfer salts, the physics boils down to that of the
Hubbard model on various lattices and at either half or one quarter
filling. We will show that the strongly correlated physics of the
various polymorphs and chemical constituents can be described within
this framework.

The Hamiltonian of the one-band Hubbard model is
\begin{eqnarray}
\hat{\cal
H}_\textrm{Hubbard}=-\sum_{ij,\sigma}(t_{ij}+\mu\delta_{ij})\hat
c_{i\sigma}^\dagger\hat c_{j\sigma} + U\sum_{i}\hat
n_{i\uparrow}\hat n_{i\downarrow},
\end{eqnarray}
where, $t_{ij}$ is the hopping integral from the site $i$ to the
site $j$, $\mu$ is the chemical potential, $U$ is the Coulomb
repulsion caused by putting two electrons on the same site, $\hat
c_{i\sigma}^{(\dagger)}$ annihilates (creates) an electron on site
$i$ with spin $\sigma$ and the number operator $\hat
n_{i\sigma}=\hat c_{i\sigma}^\dagger \hat c_{j\sigma}$.  The Hubbard
model is, perhaps, the simplest model that can describe strongly
correlated physics and is therefore an important starting point for
a complete and general description of strong correlations. Further,
the Hubbard model is also believed to describe the essential physics
of, for example, the cuprates \cite{AndersonRVB} and the cobaltates
\cite{NaxCoO}.

At some points later in this review we will also wish to discuss
extended Hubbard models. In particular we will discuss the Coulomb
repulsion when two electrons are placed on neighbouring sites, we
will refer to these as `$V$' terms and the additional term in the
Hamiltonian will be
\begin{eqnarray}
\hat{\cal H}_V=V\sum_{\langle ij\rangle}(\hat n_{i\uparrow}+\hat
n_{i\downarrow})(\hat n_{j\uparrow}+\hat n_{j\downarrow}),
\end{eqnarray}
where the angled brackets indicte the sum is over some specified set
of neighbouring sites, e.g., nearest neighbours only. Another
important term we may wish to include is the Heisenberg exchange or
`$J$' term
\begin{eqnarray}
\hat{\cal H}_J=J\sum_{\langle ij\rangle}\hat{\bf S}_i\cdot\hat{\bf
S}_j,
\end{eqnarray}
where the spin operator is $\hat{\bf S}_i=\sum_{\alpha\beta}\hat
c_{i\alpha}^\dagger \vec\sigma_{\alpha\beta} \hat c_{i\beta}$ and
$\vec\sigma=(\sigma_x,\sigma_y,\sigma_z)$ is the vector of Pauli
matrices. Lastly we may wish to include phonons and the
electron-phonon interaction in the model. The simplest approach is
to include the `Holstein' terms which treat dissipationless phononic
modes via the terms
\begin{eqnarray}
\hat{\cal H}_\textrm{Holstein}=\sum_{i\sigma\nu} g_\nu
\left(\hat{a}_{i\nu}^\dagger+\hat{a}_{i\nu}\right)\hat{n}_{i\sigma}
+ \sum_{i\nu} \omega_\nu\hat{a}_{i\nu}^\dagger\hat{a}_{i\nu},
\end{eqnarray}
where the $\hat{a}_{i\nu}^{(\dagger)}$ operators annihilate (create)
a phonon in the $\nu^\textrm{th}$ mode on the $i^\textrm{th}$
lattice site, $\omega_\nu$ is the characteristic frequency of
$\nu^\textrm{th}$ mode and $g_\nu$ is the electron-phonon coupling
between an electron on site $i$ and a phonon in the
$\nu^\textrm{th}$ mode on site $i$.

This review is \emph{not} intended to be comprehensive. Rather, we
present our views of the physics relevant to understand these
materials. This is not to say that we are the originators of all of
these ideas, but rather to stress that, as with any healthy research
field, there are a number of prominent researchers who might
disagree with the views expressed here. Space constraints will not
permit us to discuss rival theories and interpretations of data at
length, we therefore refer interested readers to some alternative
points of view \cite{alt,heat-capacity,Strack} and some more
comprehensive reviews \cite{reviews}.

\section{Minimal models}

In this review we will mostly consider materials with chemical
formulae of the form $D_2X$ where $D$ is an organic donor molecule,
for example bis\-(ethylene\-dithio)\-tetrathia\-fulvalene
[C$_{10}$S$_8$H$_8$] (often abbreviated as BEDT-TTF or ET) or
bis\-(ethylene\-dithio)tetraselena\-fulvalene [C$_{10}$Se$_8$H$_8$]
(BETS), and $X$ is an anion, e.g., I$_3$ or Cu[N(CN)$_2$]Br. The
anion accepts one electron from a pair of donor molecules which
leads, at the level of band theory, to an insulating anionic layer
and a metallic donor layer. The chemical complexity of the organic
charge transfer salts mean that few first principles calculations
have been reported \cite{DFT,Lee,Miyazaki,Kino,alpha}. However,
these suggest that a tight-binding or Huckel description of the band
structure is a reasonable approximation. In this tight binding
description a molecule serves as a `site' \cite{Kino&Fukuyama}.

\subsection{Quarter filled charge transfer salts - the $\beta''$ and $\theta$ phases}

It is clear that in the model proposed by Kino and Fukyama
\cite{Kino&Fukuyama} (described above) one expects that there should
be, on average, half a hole per site. This is indeed the case for
the $\beta''$ and $\theta$ polymorphs. These show a subtle
competition between metallic, charge ordered insulating and
superconducting phases \cite{quarter-filled,Jaime-review}. It has
been argued that this arises due to the competition of spin and
charge fluctuations. These effects have been studied in  the
extended Hubbard ($t$-$J'$-$V$-$U$)\footnote{The $J'$ indicates that
the exchange term is to next nearest neighbours. The $J$ term can be
neglected because in (or near) the charge ordered phase the are no
(few) occupied sites with occupied nearest neighbours.} model on a
square lattice \cite{quarter-filled,Jaime-review}. The importance,
for the superconducting state, of the fact that the actual lattice
has a rather lower symmetry than the square lattice has also been
stressed recently \cite{group}.

\subsection{Half-filled charge transfer salts - the $\beta$, $\beta'$, $\kappa$ and $\lambda$
phases}\label{sect:hlos}

\begin{figure}[t]
\centering
\includegraphics*[width=.7\textwidth]{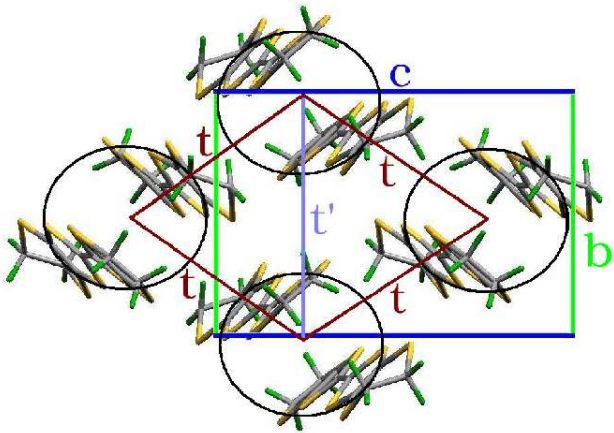}
\caption[]{A sketch illustrating why the band structure of the
half-filled layered organic charge transfer salts is approximately
that of the anisotropic triangular lattice. Black circles indicate
the dimers, the hopping integral between the two molecules in a
dimer is the largest energy scale in the problem and can be
integrated out of the effective low energy Hamiltonian leaving the
dimers as effective sites in a Hubbard model description of the
electronic degrees of freedom. The largest hopping integrals between
the dimers, $t$ and $t'$, are indicated by the maroon and blue
lines. These form an anisotropic triangular lattice. In this review
we take the $x$ and $y$ axes of the anisotropic triangular lattice
to lie along the directions of the two $t$ hopping integrals as this
is conventional in the field. This figure is based on the
crystallographic data of Rahal \etal for \NCS (a typical half-filled
organic charge transfer salt) \cite{Rahal}.}
    \label{fig:aniso}
\end{figure}

In the $\beta$, $\beta'$, $\kappa$ and $\lambda$ polymorphs there is
a single intermolecular hopping integral that is significantly
larger than the others. This led Kino and Fukuyama
\cite{Kino&Fukuyama} to propose that these two molecules behave as a
single site as the hopping between them is rapid enough that it can
be integrated out of the effective low energy theory. The two
molecules are often referred to as a `dimer'. However, it is
important to note that this does not imply that the molecules are
covalently bonded in the conventional use of the word; i.e., there
are no C-C, S-S or S-C bonds between the two molecules. Rather, the
`bond' is between the molecular orbitals themselves, this `bond'
results because of the significant overlap between the highest
occupied molecular orbitals (HOMOs) on the two molecules, and the
fact that the HOMOs are partially occupied. Thus the `bond' between
the two molecules forming the dimer is highly analogous to a
covalent bond between atoms. The dimer structure is sketched in
figure \ref{fig:aniso}, also shown are the interdimer hopping
integrals. Kino and Fukuyama gave a parameterisation of $t$ and
$t'$, the interdimer hopping integrals, in terms of the
intermolecular hopping integrals. It was later realised that this
needed to be corrected to allow the strong Coulomb repulsion between
two holes on the same molecule \cite{Ross-review}. The interdimer
hopping integrals form an anisotropic triangular lattice. Thus
geometrical frustration may be expected to play a significant role
if $t'\sim t$.

\begin{figure}[t]
\centering
\includegraphics*[width=.7\textwidth]{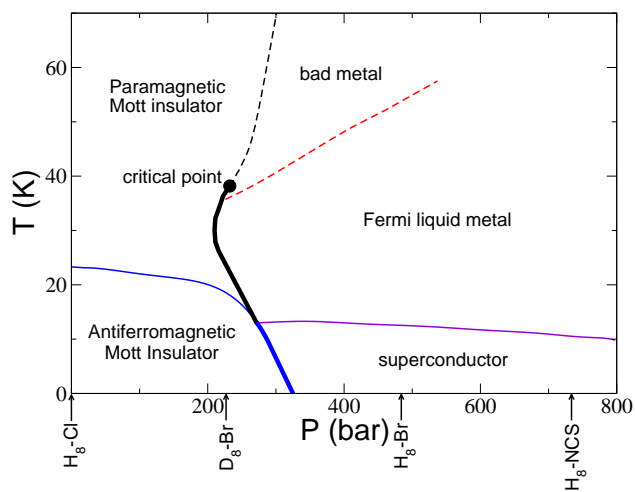}
\caption[]{A schematic pressure-temperature phase diagram of
$\kappa$-(ET)$_2X$. The first order Mott transition is shown as a
thick line, while second order transitions are thin lines and
crossovers are dashed lines. The effect `chemical' pressure for
different anions, $X$, is indicated by the arrows on the abscissa.
$P=0$ corresponds to ambient pressure for $X=$Cu[N(CN)$_2$]Cl
(abbreviated as H$_8$-Cl in the diagram). H$_8$-NCS indicates the
effective chemical pressure for $X=$Cu(NCS)$_2$ (relative to that
for $X=$Cu[N(CN)$_2$]Cl) and H$_8$-Br indicates the effective
chemical pressure for $X=$Cu[N(CN)$_2$]Br. D$_8$-Br indicates the
effective chemical pressure of $X=$Cu[N(CN)$_2$]Br with the ET
molecule fully deuterated. (Modified from \cite{RVB-organics}.)}
\label{fig:kappa-phase}
\end{figure}

The phase diagram of the $\kappa$ phase materials (the best studied
of the half-filled organic charge transfer salts) is sketched in Fig
\ref{fig:kappa-phase}. It is clear that this phase diagram cannot
result from the non-interacting model described above. However,
there is a strong Coulombic repulsion between two holes on dimer
\cite{quant-chem}. Therefore, we believe that the Hubbard model may
be the simplest model which contains the physics required to
describe these systems \cite{Kino&Fukuyama,Ross-review}. Kanoda
\cite{Kanoda} proposed that the role of pressure is to decrease
$U/W$ where $W$ is the bandwidth. As different anions lead to
different behaviours (and different unit cell volumes) in much the
same way as pressure, this can be thought of as a effective
`chemical pressure'. This idea has proven an extremely powerful
framework in which to assimilate the great deal of experimental data
now available for these systems.

An often used estimate of $U$ comes from modelling the dimer in a
two site Hubbard model with an intermolecular hopping integral,
$t_m$, and a Hubbard repulsion $U_m$ for having two holes on the
same \emph{molecule}. In the limit $U_m\gg 4t_m$ this model yields
$U\approx2t_m$ \cite{Ross-review}. However, this is rather
misleading because quantum chemistry calculations suggest that the
Coulomb repulsion between holes on neighbouring molecules in the
same dimer $V_m$ is of the order of $U_m$ \cite{quant-chem}. Thus
this term should be included and therefore the extended
($t_m$-$U_m$-$V_m$) Hubbard model should be used to estimate $U$
\cite{Greg-honours}. It is straightforward to show
\cite{Greg-honours} that for this model yields
\begin{eqnarray}
U=\frac12\left(U_m+V_m-\left[\{U_m-V_m\}^2+16t_m^2\right]^{1/2}+4t_m\right)
\label{eqn:U}
\end{eqnarray}
and in the appropriate limit, $(U_m+V_m)\gg (U_m-V_m),t_m$, we
obtain
\begin{eqnarray}
U\approx\frac12\left(U_m+V_m\right)\gg 2t_m. \label{eqn:Uapprox}
\end{eqnarray}
Taking the values of $U_m$, $V_m$ and $t_m$ reported by Fortunelli
\etal \cite{DFT} we see that equation (\ref{eqn:U}) gives
$U=3.7$~eV, this is approximated reasonably well by equation
(\ref{eqn:Uapprox}) which gives $U=4.8$~eV whereas $2t_m=0.56$~eV
which is the wrong order of magnitude. Thus we see that the plain
two site Hubbard ($t_m$-$U_m$) model significantly underestimates
$U$. A more detailed discussion of the calculation of the parameters
of the Hubbard model will be given in Sect. \ref{sect:cal-params}.
In particular that section will discuss the fact that the $U$
\emph{in vacuo} (equation \ref{eqn:U}) is significantly larger than
that in the crystal due to the polarisability of the lattice.

The majority of the remainder of this review with be devoted to
these half-filled systems. But first we briefly mention three other
classes of organic charge transfer salts.

\subsection{Half-filled charge transfer salts with magnetic anions}

An interesting class of organic charge transfer salts have been
prepared with magnetic anions. Prominent examples include
$\kappa$-(BETS)$_2$FeCl$_4$ and $\lambda$-(BETS)$_2$FeCl$_4$.
Perhaps the most novel phenomena observed in these salts is field
induced superconductivity \cite{Uji}, where superconductivity is not
observed at zero field but emerges when a large ($>15$~T) field is
applied. This occurs due to the Jaccarino-Peter effect
\cite{Jaccarino-Peter} which can be described using the
Hubbard-Kondo model \cite{Cepas}, in which a Kondo term (to describe
the magnetic anions) is added to the Hubbard model described in
Sect. \ref{sect:hlos}. For a recent review of these materials see
\cite{UjiBrooks}.

\subsection{Multiband charge transfer salts - the $\alpha$ phase}

The $\alpha$ phase salts have a complex band structure in which many
bands cross the Fermi level \cite{alpha-band}. Thus they cannot be
described in the one-band framework which is relevant to the other
organic charge transfer salts. These salts have many interesting
properties such as charge ordering, an unconventional metallic state
and superconductivity \cite{reviews}. However, space will not permit
us to discuss them at length here.

\subsection{Quasi--one-dimensional charge transfer salts}

Finally some organic charge transfer salts, e.g., the Bechgaard and
Fabre salt families, are significantly anisotropic in all three
directions and are therefore often described as
quasi--one-dimensional conductors. Typically the hopping integrals
are an order of magnitude larger in intrachain direction than in the
interchain direction, while the interchain hopping integrals are an
order of magnitude larger than those interplane \cite{Grant}. These
materials show many interesting behaviours such as spin and charge
density waves, superconductivity, possible Luttinger liquid phases,
Mott insulating states, spin-Peierls states, antiferromagnetism and
Fermi liquid behaviour \cite{Dupuis}. A key question is which of the
phenomena arise from the quasi--one-dimensionality of the system and
for which properties higher dimensional models are required to
understand the behaviour.

\vspace{12pt}

For the remainder of this review we will discuss the half-filled
organic charge transfer salts introduced in section \ref{sect:hlos}
unless it is explicitly stated otherwise.

\section{The metallic state}\label{sect:metal}

The metallic state is of fundamental importance. Firstly the organic
charge transfer salts exhibit an unconventional metallic state which
is of great interest in its own right. However, many states of
matter occur as instabilities in the metallic state: think, for
example, of the Stoner and Cooper instabilities. Thus, for example,
the difficulties in understanding the metallic state of the cuprates
have greatly compounded the difficulties in explaining the origin of
high temperature superconductivity.

Many features of the metallic states of the organic charge transfer
salts seem to agree remarkably well with the predictions of
dynamical mean field theory (DMFT). In particular, a number of
experiments show the features predicted to occur in the crossover
from a Fermi liquid to a `bad metal' described by DMFT. However,
there remain features of the metallic state that are not well
understood. For example, NMR experiments show features consistent
with a pesudogap suggesting similarities to the underdoped cuprates
(see section \ref{sect:pseudoTheory}). We will only give a very
brief discussion of DMFT itself so as to focus on the results and
the comparison with experiment. However, both a brief, accessible
introduction to DMFT \cite{Kotliar-phys-today} and a more complete
technical review  \cite{DMFT} are available in the literature.

\subsection{The bad metal and dynamical mean field theory}\label{sect:DMFT}

The low temperature metallic states of the organic charge transfer
salts are exceptionally pure Fermi liquids. This is indicated most
clearly by the observation of quantum oscillations and
angle-resolved magneto-resistance oscillations (AMRO)
\cite{Wosnitza,QuantOsc} which demonstrate that the electron
transport is coherent (at least within the planes) and that the
quasiparticle lifetime is extremely long. At low temperature the
resistivity varies like $\rho(T)=\rho_0+AT^2$
\cite{reviews,Bulaevskii,Strack,Analytis}, which is the behaviour
expected of a Fermi liquid when electron-electron interactions are
the dominant scattering mechanism \cite{Abrikosov,Baber}.

\begin{figure}[t]
\centering
\includegraphics*[width=.56\textwidth]{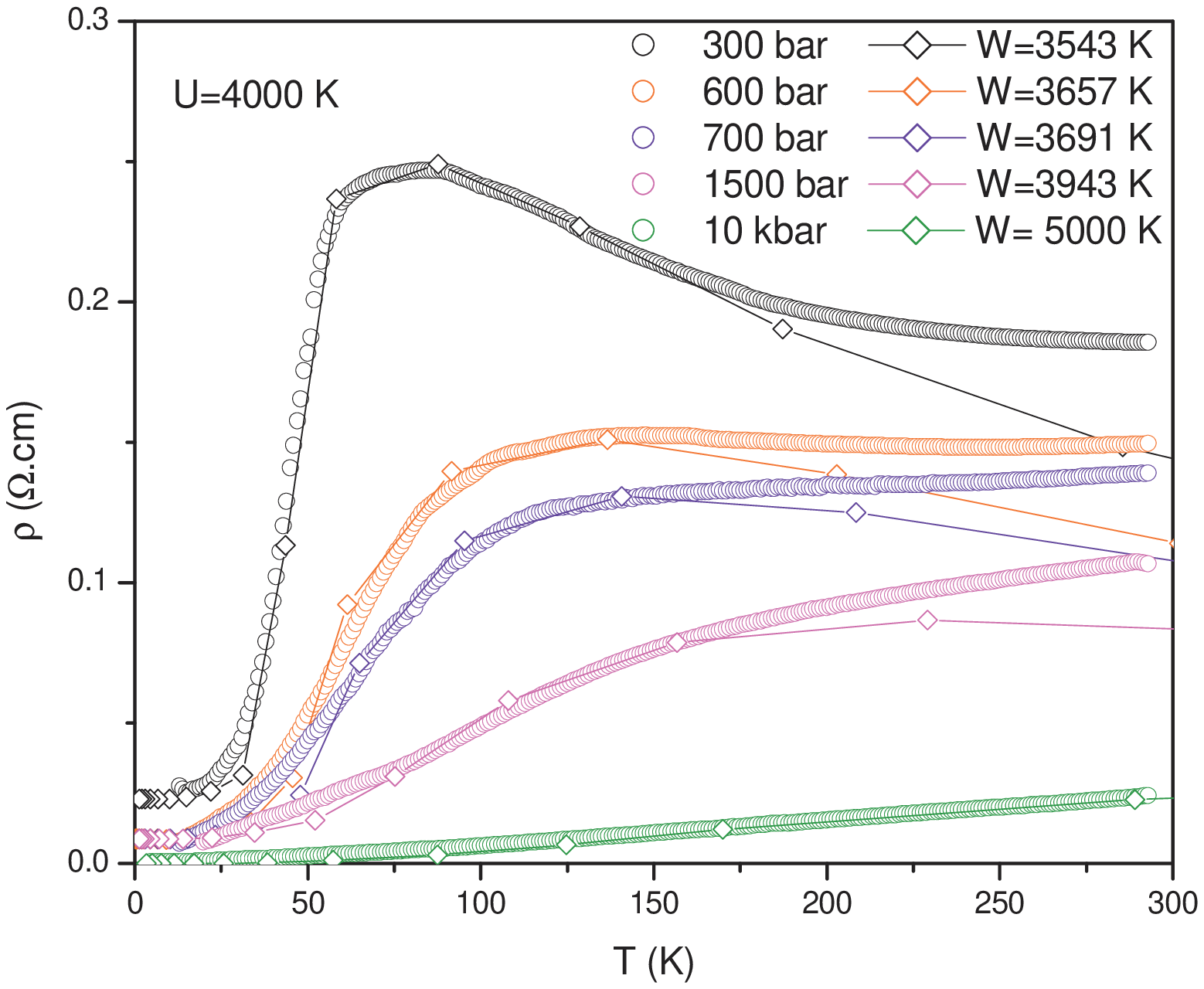}
\includegraphics*[width=.43\textwidth]{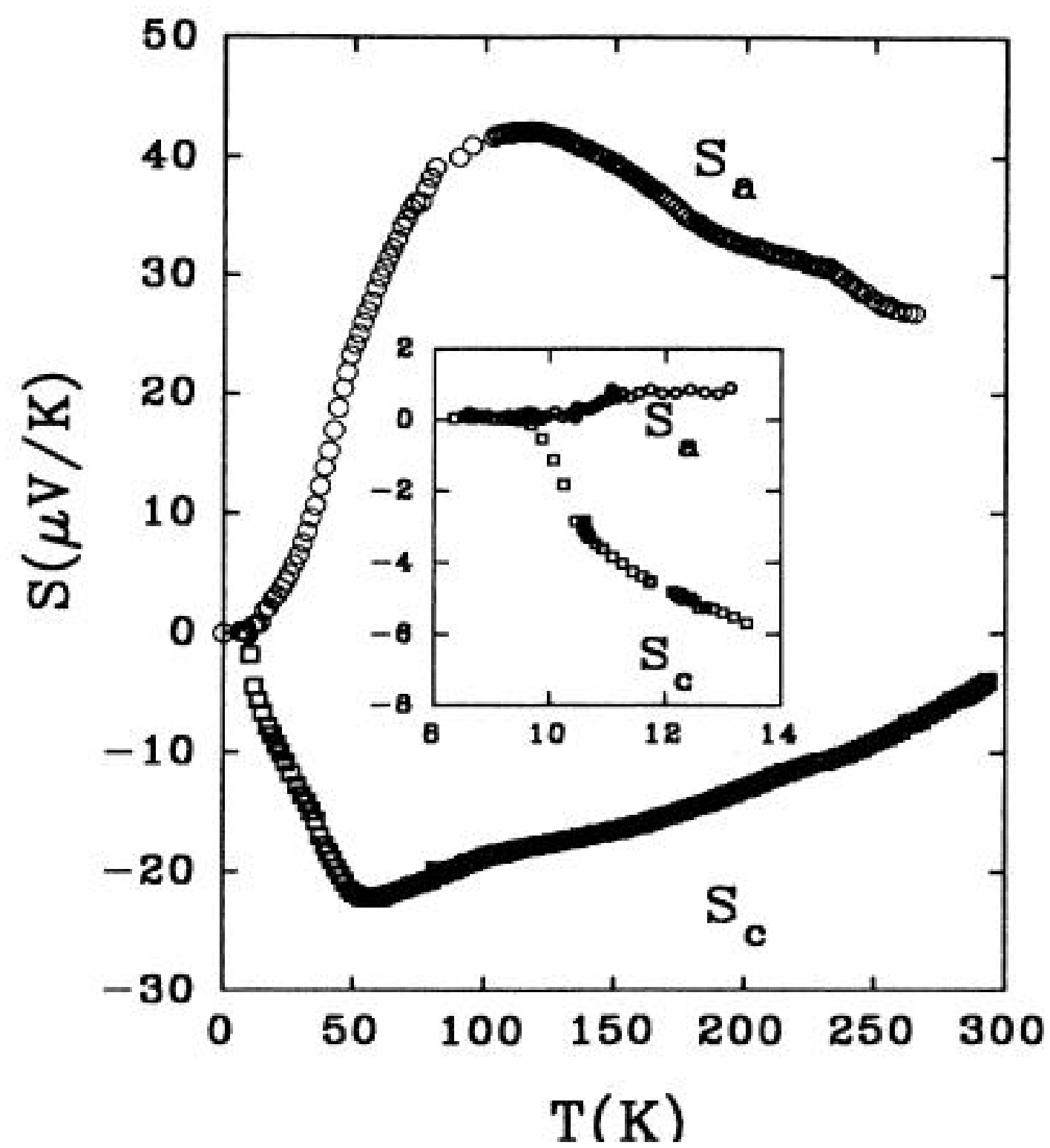}
\caption[]{Non-Fermi liquid behaviour in the resistivity and
thermopower of half-filled quasi-two-dimensional organic charge
transfer salts. The circular data points in the left panel show the
non-monotonic temperature dependence of the resistivity, $\rho$, in
\Cl (to be contrasted with the monotonic behaviour predicted by band
theory). At low pressure a broad maximum is observed in the
resistivity as is predicted by dynamical mean field theory (DMFT)
(shown as rhomboidal data points). As the pressure is increased the
broad maximum is suppressed and eventually becomes entirely absent.
DMFT reproduces these trends if pressure is taken to control the
relative interaction strength ($P\sim W/U$ \cite{Kanoda}). This data
is taken from \cite{Limelette}. The right panel shows the
temperature dependence of the thermopower, $S$, in the highly
conducting plane of \Br (taken from \cite{Yu}). The details of the
non-interacting band structure dictate that the thermopower is
positive when the heat current is parallel to the $a$-axis and
negative when the heat current is parallel to the $c$-axis. However,
along both axes the thermopower has a broad extremum at around the
same temperature as the maximum in the resistivity occurs in the
left panel. These extrema in the thermopower are predicted by DMFT
\cite{JaimeDMFT} with $U\sim W$, whereas one expects the thermopower
to be monotonic in a system with weakly interacting quasiparticles.
Thus these two experiments, which do not have a natural explanation
in a Fermi liquid picture are explained by the crossover from a `bad
metal' at high temperatures to a Fermi liquid at lower temperatures
described by DMFT.} \label{fig:Limelette} \label{fig:thermo}
\end{figure}

However, the temperature dependence of the resistivity (shown in
figure \ref{fig:Limelette}) is very different from what is expected
for simple metals [where $\rho(T)$ monotonically decreases as $T$ is
lowered]. Instead at high temperatures the resistivity
\emph{increases} as the temperature is lowered, reaching a broad
maximum, and the quadratic temperature dependence is only observed
below about $30$~K. If the mean free path is less than the lattice
spacing it is not meaningful to speak of a coherent wavevector which
describes the electronic transport. This condition allows us to
define the Mott minimum conductivity \cite{MottMin} (also known as
the Mott-Ioffe-Regel limit) which is of order $10^3$~S/cm for \NCSn.
In contrast the conductivity at the peak in the resistivity is
$\sim$1~S/cm (c.f., Fig \ref{fig:Limelette}). However, below the
peak the resistivity does decrease monotonically, thus this state is
described as a `bad metal'. Bad metal behaviour is also seen in the
alkali doped fullerides \cite{Gunnarsson}, Sr$_2$RuO$_4$
\cite{SRO4}, SrRuO$_3$ \cite{SRO3}, and VO$_2$ \cite{VO2}. All of
these systems are strongly correlated materials `near to' a Mott
transition. No Drude peak is evident in the optical conductivity of
the organic charge transfer salts above about 50~K \cite{optical}.
Further, even at low temperatures where a `Drude-like' feature does
appear, this can only be fit to the Drude form by introducing a
frequency dependencies to the scattering rate and the effective mass
\cite{optical}. On the other hand a broad, high frequency peak is
observed at all temperatures. This is suppressed somewhat at low
temperatures.

The DMFT of the Hubbard model provides a description of all of this
physics \cite{JaimeDMFT,Limelette}. DMFT works by mapping the
Hubbard model onto the Anderson model for a single magnetic impurity
in a bath of conduction electrons \cite{DMFT,Kotliar-phys-today}.
This procedure is exact in the limits of infinite spatial dimensions
or infinite lattice coordination number. For half-filled systems
close to a Mott transition DMFT predicts a Fermi liquid at low
temperatures, and a crossover to incoherent transport as the
temperature is raised. The crossover temperature scale, $T_{coh}$,
is related to the destruction of Kondo screening and Fermi liquid
behaviour with increasing temperature (above the Kondo temperature
$T_K$) in the Anderson model. In the Anderson model the conduction
electrons are strongly scattered by a (localised) magnetic impurity
for $T>T_K$. But for $T<T_K$ a quantum coherent singlet forms
between the impurity and the conduction electrons \cite{Hewson}. In
the DMFT of the Hubbard model for $T>T_{coh}$ the electrons are
quasi-localised and the electrons on the single site treated exactly
strongly scatter those in the bath. However, for $T<T_{coh}$
transport is coherent and the electrons only scatter one another
weakly, thus Fermi liquid behaviour is regained. This is why, for
example, the temperature dependence of the resistivities of the
Anderson and Hubbard models are so similar \cite{Hewson,DMFT}. DMFT
predicts that there is no Drude peak in the bad metal phase and that
most of the spectral weight is contained in high energy features.
This is because much of the spectral weight is transferred to the
`Hubbard bands' that will emerge in the Mott insulating state
\cite{DMFT,Kotliar-phys-today}. This prediction is clearly in good
agrement with the optical conductivity measurements \cite{optical}
described above.

The success of DMFT in describing the transport properties and the
phase diagram of many organic charge transfer salts down to
temperatures of about 10~K (where, for example, superconductivity
becomes important) has been rather puzzling, until recently
\cite{NaxCoO}, given that these materials are quasi--two-dimensional
and that DMFT is only expected to be a good approximation in the
limit of high spatial dimension or high co-ordination number.
However, the applicability of DMFT to low-dimensional systems with
large frustration is consistent with the fact that for frustrated
magnetic models a Curie-Weiss law holds down to a much lower
temperature than for unfrustrated models \cite{Ramirez,tri-series}.
Deviations from Curie-Weiss behaviour result from spatially
dependent correlations. Hence, we expect that a DMFT treatment of
the Hubbard model on frustrated lattices will be a good
approximation down to much lower temperatures than it is for
unfrustrated models.

However, DMFT is not simply a theory for describing the
conductivity: good agrement with experiment is found for many other
properties. For example, measurements of the linear coefficient of
the specific heat, $\gamma$, of \Br and \NCS \cite{Cv} show that the
effective mass, $m^*$, is several (3-5) times that predicted by band
theory \cite{Huckel}. Similar effective masses are found in quantum
oscillation experiments \cite{QuantOsc,Wosnitza}. DMFT predicts this
large mass enhancement. This ia a result of the strong electronic
correlations.

Experimentally, the thermopower, shown in figure \ref{fig:thermo},
in the highly conducting ($ac$) plane of \Br is positive (negative)
in the $a$ ($c$) direction.\footnote{The sign difference results
from the details of the band structure of \Br and corresponds to
hole (electron) like conduction along the $a$ ($c$) directions
\cite{Yu}.} The thermopower, $S$, shows a broad maximum (minimum) at
around 100~K (60~K) \cite{Yu}, i.e., at about the same temperature
the resistivity shows a maximum (figure \ref{fig:Limelette}). The
thermopower predicted for a weakly interacting metallic state with
the band structure of \Br is $\sim$5 times smaller than that
observed experimentally \cite{Yu,Ross-review}. DMFT predicts an
extremum in the thermopower at $T\sim T_{coh}$ for large $U/W$
\cite{JaimeDMFT} and the large effective mass predicted by DMFT
gives rise to the large thermopower as $S/T\propto m^*$, in good
agrement with experiment.

A strong decrease in the ultrasonic velocity is observed at
temperatures $\sim$40~K in \Br \cite{Frikach}, \NCS \cite{Frikach}
and \Cl (under pressure \cite{Fournier}). The pressure (and chemical
pressure) dependence of the temperature at which these anomalies are
observed show that they occur at the same temperature as the other
anomalies that, we have argued above, correspond to $T_{coh}$. Does
this then suggest that phonons play an important role in this
physics? DMFT shows that this is not the case. DMFT studies of the
Hubbard-Holstein model \cite{Jaime-HH-DMFT,Georges-HH-DMFT} in the
limit where the electron-electron interactions are much stronger
than the electron-phonon interactions ($U\gg g$) predict phonon
anomalies at $T_{coh}$. However, these anomalies are parasitic: the
changes in the behaviour of the electrons cause the change in the
behaviour of the lattice vibrations because of the electron-phonon
coupling, and the phonons do not play a significant role in driving
the crossover from the Fermi liquid to the `bad metal'.

The Sommerfeld--Wilson ratio is defined as
$R_W={[\chi(T=0)\gamma_0]/(\gamma\chi_0)}$,
where $\chi(T)$ is the magnetic susceptibility, $\chi_0$ is the zero
temperature magnetic susceptibility of the non-interacting electron
gas and $\gamma_0$ is the linear coefficient of the heat capacity
for the non-interacting electron gas. Clearly for the
non-interacting system $R_W=1$, but the Kondo model predicts
$R_W=2$. Therefore, as DMFT exploits the deep connections between
Kondo physics and the Mott-Hubbard transition DMFT predicts $R_W>1$
in the organics. Experimentally $R_W$ is $1.5\pm0.2$ in both \Br and
\NCS \cite{Ross-Wilson}.

The Kadowaki-Woods ratio is $R_{KW}=A/\gamma^2$, where $A$ is the
quadratic coefficient of the resistivity. In many strongly
correlated materials
$R_{KW}\sim10.5~\mu\Omega$~cm~K$^2$~mol$^2$~J$^{-2}$. The
Kadowaki-Woods ratio is significantly larger than this in the
half-filled layered organic materials, even when differences in the
unit cell volume are allowed for \cite{Hussey,Jacko}. This has led
to suggestions that the quadratic temperature dependence of the
resistivity may not result from strong electronic correlations, but
from phonons \cite{Strack,T2phonons}. However, one must be careful
to allow for the fact that the only reliable resistivity
measurements in the organic charge transfer salts are perpendicular
to the highly conducting plane. When this is allowed for the
observed Kadowaki-Woods ratio is the expected order of magnitude
\cite{Jacko}.

Collectively the experiments described above (and those studying the
Mott transition which we review in section \ref{sect:Mott}) show
that a weakly interacting, Fermi liquid, description is not
sufficient to explain the full temperature dependence of the
thermodynamic and transport properties of the half-filled organic
charge transfer salts. However, DMFT, which includes the effects of
strong electron-electron interactions, and reproduces Fermi liquid
theory below $T_{coh}$, can provide both a quantitative
\cite{Limelette} and qualitative \cite{JaimeDMFT} description of the
full temperature dependence. However, DMFT is a purely local theory.
Therefore, any properties with a significant $\bf k$-dependence will
not be properly described by DMFT. We will discuss some such
possible features in section \ref{sect:pseudoTheory}.

\section{The Mott transition}\label{sect:Mott}

DMFT also provides a description of the, first order, Mott
metal-insulator transition \cite{Kotliar-phys-today,DMFT}. For $U=0$
the density of states (DOS) calculated from DMFT is, as one should
expect, that of the tight-binding model. For a weakly interacting
system the DOS will be only weakly renormalised by the interactions.
However, as $U$ is turned on the system becomes strongly interacting
and a peak in the DOS will emerge at the Fermi energy. This peak is
associated with quasiparticles. The total spectral weight associated
with the quasiparticles is $Z=m_b/m^*<1$, where $m_b$ is the band
mass of the electron, hence $Z$ is called the quasiparticle weight.
The remainder of the spectral weight is transferred to two broad
bands centred on $\pm U/2$ of width $\sim W$, known as the Hubbard
bands. The Hubbard bands correspond to quasi-localised states with
the lower Hubbard band (centred at $-U/2$) corresponding to singly
occupied sites and the upper Hubbard band (centred at $+U/2$)
corresponding to doubly occupied sites. As $U$ is further increased
the quasiparticle peak narrows ($Z$ decreases) and more spectral
weight is transferred to Hubbard bands. When $U$ is increased above
the critical $U\sim W$ for the Mott transition the quasiparticle
peak vanishes and all of the spectral weight resides in the Hubbard
bands. There is now no density of states at the Fermi energy and so
we have an insulator. A number of studies have found that this
transition is first order within the DMFT framework
\cite{DMFT,first-order}.

It is conceptually useful to compare the Mott transition in the
organic charge transfer salts (figure \ref{fig:kappa-phase}) and
that predicted for the Hubbard model on the anisotropic triangular
lattice by DMFT with the liquid-gas transition. The difference
between the liquid and the gas is their densities. One may move from
the gas to the liquid in either of two ways: either directly, by
increasing the pressure through a first order phase transition where
the density changes discontinuously; or by first increasing the
temperature above the critical temperature and then passing around
the critical point. If one passes around the critical point no phase
transition is observed and the density varies continuously through
the fluid phase. The Mott transition behaves in much the same way.
In the Mott insulator the electrons are localised and hence the
conductivity is poor whereas the metal has a large conductivity. We
may either drive the Mott transition by increasing the pressure and
passing directly through the first order transition where the degree
of localisation changes discontinuously, or else we may first
increase the temperature above the critical temperature where the
line of first order transitions ends (see figure
\ref{fig:kappa-phase}) and thus pass continuously from the
insulating phase to the metallic phase. In this picture the
`bad-metal' is somewhat analogous to the fluid. As we move through
the `bad-metal' the conductivity changes continuously as we pass
from the localised behaviour of the insulator to the coherent
excitations characteristic of the Fermi liquid. Although one should
note that the analogy is not exact as the `bad metal' regime extends
slightly below the critical point (c.f. figure
\ref{fig:kappa-phase}).

\subsection{The critical point}

We now turn our discussion to the critical point itself.
Unsurprisingly, DMFT predicts mean-field critical exponents (see
table \ref{tab:exponents}). Limelette \etal \cite{LimeletteVO}
measured the critical exponents of the Mott transition in V$_2$O$_3$
\cite{LimeletteVO}.\footnote{These exponents can be measured by
studying the conductivity near the Mott transition. The conductivity
gives access to the critical behaviour as it is related to the
thermodynamic ground state via linear response theory. The critical
exponents can be determined from the conductivity, $\sigma$, as:
$(\sigma-\sigma_c)\sim(T_c-T)^{\beta}$,
$(d\sigma/dP)_T\sim|T-T_c|^{-\gamma}$ and
$(\sigma-\sigma_c)\sim(P-P_c)^{1/\delta}$, where $\sigma_c$ is the
conductivity at the critical point, $T$ is the temperature, $T_c$ is
the critical temperature, $P$ is the pressure and $P_c$ is the
critical pressure. To interpret the order parameter exponent on the
critical isotherm, $\delta$, one should recall that the order
parameter of the Mott transition is the half-width of the
coexistence curve, which vanishes at the critical point. This is
entirely analogous to the liquid-gas transition \cite{Goldenfeld}.
It is often stated that the order parameter of the liquid-gas
transition is the difference in the densities of the liquid and the
gas. This statement needs to be interpreted carefully. Strictly the
density difference between the liquid and the gas phase \emph{when
they are in coexistence} is a good order parameter. This is, of
course, proportional to the half-width of the coexistence curve
because of the Maxwell equal area construction. However, the density
difference when the liquid and gas when they are not in coexistence
is \emph{not} a good order parameter. These comments are equally
valid for the difference in conductivities between the metal and
insulator in the Mott transition.} They observed three dimensional
(3D) Ising exponents (the same as are seen in the liquid-gas
transition), but only in a extremely narrow critical region around
the critical point ($P/P_c,\; T/T_c\lesssim10^{-2}$). Limelette
\etal observed mean-field critical exponents except in this
extremely small critical region near the critical point which, they
argued, implies that there are bound pairs of doubly occupied and
vacant sites near the Mott transition. These bound states can exist
over large length scales (of order nm) and are argued to be the root
cause of the success of DMFT. This is analogous to the success of
the (mean field) BCS theory of superconductors which results from
the large coherence length (and hence small critical region) in
conventional superconductors.

However, recently Kagawa \etal \cite{Kagawa} have found that the
critical exponents for the critical end point of the Mott transition
in \Cl do not correspond to either the Ising or mean field
universality classes. Indeed the critical exponents do not belong to
any universality class that had previously been observed (see table
\ref{tab:exponents}). Crucially, in spite of being very different
from those for the Ising, XY or Heisenberg universality classes, the
observed exponents obey the standard scaling relation
$(\delta-1)\beta=\gamma$, within experimental error.

\begin{table}[t]
\begin{tabular}{|l|c|c|c|c|c|}
  \hline
  & Example system & \hspace{0.5cm} $\beta$ \hspace{0.5cm} & \hspace{0.5cm} $\gamma$ \hspace{0.5cm} & \hspace{0.5cm} $\delta$ \hspace{0.5cm} \\
  \hline
  mean field  \cite{Goldenfeld} & DMFT &  1/2 & 1 & 3 \\
  3D Ising \cite{Goldenfeld} & Liquid-gas &  0.33 & 1.24 & 4.8 \\
  3D XY  & Superconductor &  0.35 & 1.3 & 4.8 \\
  3D Heisenberg & Ferromagnet &  0.36 & 1.4 & 4.8 \\
  2D Ising  \cite{Goldenfeld} & Physisorption &  1/8 & 7/4 & 15 \\
  mean field at MQCP & MQCP \cite{Imada,Misawa} &  1 & 1 & 2 \\
  mean field near MQCP & MQCP \cite{Misawa} &  $1\rightarrow0.33$ & $1\rightarrow1.24$ & $2\rightarrow4.8$ \\
  3D metal-insulator &  V$_2$O$_3$\cite{LimeletteVO} &  $0.34\rightarrow0.5$ & 1 & $5\rightarrow3$ \\
  2D metal-insulator? &  \Cl \cite{Kagawa} &  0.9-1 & 0.9-1 & 1.9-2 \\
  \hline
\end{tabular}
\caption{Comparison of the critical exponents observed for some
common universality classes with those seen at the critical point of
the Mott transition in \Cln. The symbol $X\rightarrow Y$ indicates
that the measured exponent crosses over from $X$ in a small critical
region near the critical point, to $Y$ in a larger region further
from the critical point.}\label{tab:exponents}
\end{table}

Remarkably, Imada \cite{Imada} had predicted these exponents from
phenomenological theories of the Mott transition. Misawa \etal
\cite{Misawa} have now shown that these critical exponents may be
derived within the Hartree-Fock approximation from the Hubbard model
on the anisotropic triangular lattice. Within the Hartree-Fock
approximation the metal becomes magnetically ordered at relatively
small $U/W$. But the metal insulator transition does not occur until
$U\sim W$. Thus there is no symmetry breaking at the metal-insulator
transition within the Hartree-Fock approximation. Misawa \etal find
that for $0.056\approx t_{c1}'/t<t'/t<t_{c2}'/t\approx0.365$ a line
of first order phase transitions ends at a finite temperature
critical point. At $t_{c1}'$ and $t_{c2}'$ this critical point is
driven to $T=0$ and for $t'/t<t_{c1}'/t$ or $t'/t>t_{c2}'/t$ a
quantum critical point is observed. $t'=t_{c1}'$ and $t'=t_{c2}'$
are therefore termed marginal quantum critical points (MQCPs). At
the MQPCs Misawa \etal find that the critical exponents are
$\beta=1$, $\gamma=1$ and $\delta=2$, which satisfy the scaling
relation $(\delta-1)\beta=\gamma$, and agree with those measured by
Kagawa \etal \cite{Kagawa} in \Cl within experimental error.
However, one should recall that these exponents are only strictly
valid at $T=0$ and the critical point in \Cl occurs at 39.7~K -
where we might expect to regain the standard Ising critical
exponents. However, Misawa \etal stress that the true critical
exponents will only apply within a small critical region very close
to the phase transition and that beyond that region the observed
exponents will crossover to those characteristic of the MQCP. Misawa
\etal estimate that experimental resolution of Kagawa \etal is two
orders of magnitude worse than that required to see the Ising
critical behaviour. Misawa \etal stress that the fact that magnetic
order exists on both the insulating and metallic sides of the phase
transition is vital for their theory.\footnote{Note that Imada's
\cite{Imada} phenomenological theories which predict the same
exponents do so in the case of the non-magnetic--insulator to
non-magnetic--metal transition.} We should note that that in \Cl
antiferromagnetic ordering does not occur over the entirety of the
Mott insulating phase, but only at low temperatures (see figure
\ref{fig:kappa-phase}). Therefore, the Mott transition in \Cl is not
accompanied by any symmetry breaking in the vicinity of the critical
point. Importantly, the Ginzburg criterion shows that the upper
critical dimension is two and so the mean field Hartree-Fock
treatment, which neglects all fluctuations, appears likely to be
correct in cases where no symmetry is broken at the critical point.
It is important to note the role that frustration plays in this
scenario as for weakly frustrated lattices ($t'/t<t_{c1}'/t$ or
$t'/t>t_{c2}'/t$) these unconventional critical exponents are not
predicted.

\section{The superconducting state}\label{sect:sc}

The organic charge transfer salts are often referred to as the
organic superconductors. However, this is somewhat misleading both
because of the range of other phenomena observed in the organic
charge transfer salts and because of the range of other materials
which fit the description `organic superconductors'. Other organic
superconductors include the alkali doped fullerides
\cite{Gunnarsson}, intercalated graphite \cite{graphite} and ion
implanted polymers \cite{Mico}. However, below we discuss only
superconductivity in the quasi-two-dimensional half-filled organic
charge transfer salts.

The phrase `unconventional superconductivity' has several meanings.
These meanings are best illustrated by comparison to elemental
superconductors, which are described by BCS theory \cite{BCS} (or
more accurately by Eliashberg theory \cite{Eliashberg} or the
density functional theory of superconductivity \cite{SCDFT}). In
these theories the electrons form Cooper pairs due to an effective
attractive interaction between electrons mediated by the exchange of
phonons. The BCS state breaks gauge symmetry but no other symmetry
of the crystal. Thus the phrase `unconventional superconductivity'
can be applied to any superconductor that does not satisfy any one
of these three requirements (BCS, phononic pairing mechanism and no
additional symmetry breaking) for conventional superconductivity. In
this section we review the experimental evidence which shows that
the superconductivity in the half-filled organic charge transfer
salts is unconventional in all three of the senses described above.
We delay a detailed discussion of the theory until section
\ref{sect:SCtheory}.

\subsection{Pairing symmetry (evidence for additional symmetry
breaking)}\label{sect:symetry}

In the original formulation of BCS theory \cite{BCS} one assumes
that the effective attractive pairwise interaction responsible for
superconductivity, $V$, is spatially uniform. This means that the
superconducting order parameter $\Delta$ is also isotropic as
$\Delta=\sum_{\bf k}V\langle c_{{\bf k}\uparrow} c_{-{\bf
k}\downarrow}\rangle$. This assumption is somewhat unphysical, but
it is straightforward to generalise BCS theory to allow for momentum
dependent effective interactions, $V_{\bf k}$ \cite{Ketterson&Song}.
In this case we should also allow the order parameter to develop a
${\bf k}$-dependence by defining $\Delta_{\bf k}=\sum_{{\bf
k}'}V_{{\bf k-k}'}\langle c_{{\bf k}'\uparrow} c_{-{\bf
k'}\downarrow}\rangle$. It is now natural to ask what symmetry
$\Delta_{\bf k}$ has. Note that although we have introduced these
ideas in the context of BCS theory the symmetry arguments that
follow are based basic truths about quantum mechanics and do not
depend on the details of the microscopic theory of
superconductivity. Thus symmetry analyses are a powerful framework
in which to understand the phenomenology of unconventional
superconductors \cite{group,James_adv_phys,Sigrist&Ueda}.

\subsubsection{A brief introduction to the symmetry of pairing
states}

A fundamental theorem of quantum mechanics is that the eigenstates
must transform according to an irreducible representation of the
symmetry group of the Hamiltonian
\cite{Tinkham-group}.\footnote{Here we neglect the possibility of
accidental degeneracies as they complicate the discussion
considerably, but do not really change the physics as the
accidentally degenerate states can always be decomposed into a set
of eigenstates that transform under the operations of the group in
the same way as a set of the irreducible representations of the
symmetry group of the Hamiltonian.} This is true regardless of the
complexity of the quantum many-body Hamiltonian. Therefore, one
expects on very general grounds that the superconducting order
parameter (which has the same symmetry as the wavefunction
describing the superconducting state\footnote{Assuming no other
phase transition accompanies the superconducting transition.}) will
have the symmetry of a particular irreducible representation of the
group describing the symmetry of the normal state (and hence the
Hamiltonian at $T_c$). This argument can be made rigourous exactly
at the critical temperature, and may be expected to hold below $T_c$
provided there are no further phase transitions.

In elemental superconductors it is found that, although $\Delta_{\bf
k}$ is not completely isotropic \cite{Morse} the order parameter
does have the same symmetry as the crystal. The first material found
where this did not appear to  be the case was in superfluid $^3$He.
Clearly there is no crystal in liquid $^3$He and so the normal state
has the full symmetry of free space. Therefore, the expectation that
the order parameter will transform like a particular representation
of the group describing the symmetry of the normal state can be
recast as the claim that if we expand the order parameter in terms
of the spherical harmonics, $Y_{lm}({\bf {\hat k}})$,
\begin{eqnarray}
\Delta_{{\bf k}}=\sum_{l=0}^\infty \sum_{m=-l}^l \eta_{lm}
Y_{lm}({\bf {\hat k}}),
\end{eqnarray}
then we will find that we only require $\eta_{lm}$ to be finite for
one particular value of $l$. Thus we find that
\begin{eqnarray}
\Delta_{{\bf k}}= \sum_{m=-l}^l \eta_{lm} Y_{lm}({\bf {\hat k}}).
\end{eqnarray}
It is natural to refer to superconductors in which the non-zero
$\eta_{lm}$ are $l=0$ as $s$-wave, superconductors in which the
finite $\eta_{lm}$ are $l=1$ as $p$-wave, superconductors in which
the finite $\eta_{lm}$ are $l=2$ as $d$-wave, and so on, by analogy
with atomic physics. It has been established that $^3$He is a
$p$-wave superconductor \cite{Leggett}.

Historically, the discovery that superfluid $^3$He has a lower
symmetry than normal $^3$He proceeded any analogous discoveries in
superconductors by some time. However, there are now several
materials in which the superconducting state is believed to have a
lower symmetry than the normal state. These include, \SROn, the
cuprates, several heavy fermion materials and, we will argue below
the layered organic charge transfer salts. One often hears
statements such as ``the cuprates are $d$-wave superconductors''. It
is not immediately clear what this means as the normal state of the
cuprates does not have the full symmetry of free space (due to the
crystal lattice) and so the spherical harmonics are not an
appropriate basis in which to expand the order parameter.

In a crystal the natural basis is that of the irreducible
representations of the point group of the crystal.\footnote{We
assume throughout that the superconducting state does not break
translational symmetry.} (Table \ref{tab:D4h} describes the point
group $D_{4h}$ which is that of most cuprates and a number of other
unconventional superconductors and tables
\ref{tab:D2h}-\ref{tab:C2v} give three examples of point groups
relevant to the organic charge transfer salts.) In this case the
order parameter may be written as
\begin{eqnarray}
\Delta_{{\bf k}}= \sum_{i=0}^{d^\Gamma} \eta_{i} \psi_i^\Gamma({\bf
k}),
\end{eqnarray}
where $\psi_i^\Gamma({\bf k})$ are the basis functions of the
$\Gamma^\textrm{th}$ irreducible representation and $d^\Gamma$ is
the dimension of $\Gamma$.

The above discussion is only valid for singlet superconductors.
Triplet superconductivity does not greatly complicate this analysis,
but one does have to include a sum over the three spin projections
of the $S=1$ Cooper pairs and generalise the order parameter
slightly (see \cite{Sigrist&Ueda}, \cite{James_adv_phys} or
\cite{Vollhardt} for a careful discussion of the group theoretic
classification of triplet states).

\begin{table}[t]
\begin{tabular}{|l|cccccccccc|c|c|c|c|}
  \hline
  Irrep & $E$ & $C_4^c$ & $C_2^{a/b}$ & $C_2^d$ & $C_2^c$ & $i$ & $S_4^c$ & $\sigma^\frac{ac}{bc}$ & $\sigma^{dc}$ & $\sigma^{ab}$
  & \begin{tabular}{c}
    Required \\
    nodes
  \end{tabular}
  & \begin{tabular}{c}
    Example  \\
    basis functions
  \end{tabular}
  & \begin{tabular}{c}
    Colloquial \\
    names
  \end{tabular}\\
  \hline
  $A_{1g}$ & 1 &  1 &  1 &  1 &  1 &  1 &  1 &  1 &  1 &  1 & none & $1_{\bf k}$, $X_{\bf k}^2+Y_{\bf k}^2$, $Z_{\bf k}^2$ & $s$ \\
  $A_{2g}$ & 1 &  1 & -1 & -1 &  1 &  1 &  1 &  1 & -1 & -1 & line & $X_{\bf k}Y_{\bf k}(X_{\bf k}^2-Y_{\bf k}^2)$ & $g$ \\
  $B_{1g}$ & 1 & -1 & -1 &  1 &  1 &  1 &  1 & -1 &  1 & -1 & line & $X_{\bf k}^2-Y_{\bf k}^2$ & $d_{x^2-y^2}$ \\
  $B_{2g}$ & 1 & -1 &  1 & -1 &  1 &  1 &  1 & -1 & -1 &  1 & line & $X_{\bf k}Y_{\bf k}$ & $d_{xy}$ \\
  $E_{g}$  & 2 & -2 &  0 &  0 &  0 &  2 & -2 &  0 &  0 &  0 & none & $(X_{\bf k}Z_{\bf k},Y_{\bf k}Z_{\bf k})$ & $(d_{xz},d_{yz})$ \\
  $A_{1u}$ & 1 &  1 &  1 &  1 &  1 & -1 & -1 & -1 & -1 & -1 & none & $X_{\bf k}^2Y_{\bf k}^2Z_{\bf k}$ & $h$ \\
  $A_{2u}$ & 1 &  1 & -1 & -1 &  1 & -1 & -1 & -1 &  1 &  1 & line & $Z_{\bf k}$ & $p_z$ \\
  $B_{1u}$ & 1 & -1 & -1 &  1 &  1 & -1 & -1 &  1 & -1 &  1 & line & $X_{\bf k}Y_{\bf k}Z_{\bf k}$ & $f_{xyz}$ \\
  $B_{2u}$ & 1 & -1 &  1 & -1 &  1 & -1 & -1 &  1 &  1 & -1 & line & $(X_{\bf k}^2-Y_{\bf k}^2)Z_{\bf k}$ & $f_{(x^2-y^2)z}$ \\
  $E_{u}$  & 2 & -2 &  0 &  0 &  0 & -2 &  2 &  0 &  0 &  0 & none & $(X_{\bf k},Y_{\bf k})$ & $(p_x,p_y)$ \\
  \hline
\end{tabular}
\caption{The character table, symmetry required nodes in the
superconducting energy gap and some basis functions of the
irreducible representations of the point group $D_{4h}$. This is the
point group of many unconventional superconductors with tetragonal
crystals, including many of the cuprates and \SROn. We assume that
the $x$, $y$ and $z$ axes are, respectively, parallel to the $a$,
$b$ and $c$ axes. The functions $1_{\bf k}$, $X_{\bf k}$, $Y_{\bf
k}$ and $Z_{\bf k}$ may be any functions which transform,
respectively, as 1, $k_x$, $k_y$ and $k_z$ under the operations of
the group and satisfy translational symmetry. The operations of the
group are the identity ($E$), rotation by $\pi$ and $\pi/2$ about
the $c$ axis (respectively $C_2^c$ and $C_4^c$), rotation by $\pi$
about the $a$ and $b$ axes (which have the same characters, which we
therefore label $C_{2}^{a/b}$), rotation by $\pi$ about the either
of the diagonal of the $a$-$b$ plane (which have the same characters
and are labelled $C_{2}^d$), inversion ($i$: inversion symmetry
takes $\bf k$ to $-\bf k$), an improper by $\pi/4$ about the $c$
axes ($S_4^c$: an improper is a rotation followed by a reflection
through the plane perpendicular to the axis of rotation), reflection
through the $ab$ plane ($\sigma^{ab}$) reflection through the $ac$
and $bc$ planes (which have the same characters, which we therefore
label $\sigma^\frac{ac}{bc}$) and reflection through the planes
specified by $(x+y)z=0$ and $(x-y)z=0$ (which have the same
characters and are labelled $\sigma^{dc}$). A brief explanation of
characters is given in the caption to table
\ref{tab:D2h}.}\label{tab:D4h}
\end{table}

Thus when the statement is made that ``the cuprates are
$d_{x^2-y^2}$ superconductors'', what is meant is that the
superconducting state transforms like the $B_{1g}$ representation of
the group $D_{4h}$ (c.f. table \ref{tab:D4h}). However, all possible
basis functions of the $B_{1g}$ representation of $D_{4h}$ vanish
along the lines $k_x=\pm k_y$. Thus the nodes of the order parameter
(which are often all one is concerned with, see below and section
\ref{sect:probes}) are the same as for the $d_{x^2-y^2}$ spherical
harmonic. Thus in analogy with the case of $^3$He, people often
refer to `$s$-wave', `$p$-wave' or `$d$-wave' superconductors. One
should always bare in mind what this naming conventional really
means, as confusion can arise (and repeatedly has arisen) when the
terms are used carelessly. However, this terminology is all but
universal and therefore we will make use of it, but we will use
inverted commas to remind the reader that an analogy is being drawn
and that we really mean a particular irreducible representation of
the point group of the crystal in question.

\subsubsection{Singlet or triplet?}

As all of the half-filled organic charge transfer salts have
inversion symmetry singlet and triplet states are distinct
\cite{James_adv_phys}. Thus the first question we should ask is
whether the superconducting states of these materials are singlet or
triplet. Measurements of the $^{13}$C NMR Knight shift
\cite{de_Soto,Mayaffre,NMRreview} in \Brn, with the magnetic field,
${\bf H}$, parallel to the conducting planes, show that as $T
\rightarrow 0$ so does the Knight shift. This single experiment does
not actually rule out triplet pairing, although it does make triplet
pairing extremely unlikely. This experiment is compatible with a
triplet state in which ${\bf d}({\bf k})\times{\bf H} = 0$ where
${\bf d}({\bf k})$ is the usual Balian--Werthamer order parameter
for triplet superconductivity \cite{Balian&Werthamer,Vollhardt}.
However, Zuo \etal \cite{Zuo} measured the critical field as a
function of temperature with ${\bf H}$ parallel to the conducting
planes. In this configuration no orbital currents flow so the
critical field is due to Clogston--Chandrasekhar (or Pauli) limit
\cite{Clogston,Chandrasekhar,Ben3}. There is no
Clogston--Chandrasekhar limit for ${\bf H} \perp c$ for triplet
states compatible with measured Knight shift. Thus, for such states
there would be no critical field with ${\bf H} \| b$ (in fact for
such states one would increase $T_c$ by applying a field parallel to
the b-axis \cite{Ben3}). Experimentally \cite{Murata87} it is found
that superconductivity is destroyed by a magnetic field parallel to
the b-axis. Therefore, only when considered together do the three
experiments discussed above \cite{de_Soto,Zuo,Murata87} strictly
rule out triplet pairing. 
Further evidence
for Clogston--Chandrasekhar limiting, and hence singlet pairing,
comes from the observation that the in plane upper critical field is
independent of the field direction \cite{Nam}. Given the anisotropic
nature of the Fermi surface of \Br it is extremely unlikely that
orbital mechanisms for the destruction of superconductivity would be
so isotropic.

\begin{table}[t]
\begin{tabular}{|l|cccccccc|c|c|c|c|}
  \hline
  Irrep & $E$ & $C_2^a$ & $C_2^b$ & $C_2^c$ & $i$ & $\sigma^{bc}$ & $\sigma^{ac}$ & $\sigma^{ab}$
  & \begin{tabular}{c}
    Required \\
    nodes
  \end{tabular}
& Example basis functions
  & \begin{tabular}{c}
    Colloquial \\
    names
  \end{tabular}\\
  \hline
  $A_{1g}$ & 1 & 1 & 1 & 1 & 1 & 1 & 1 & 1 & none & $1_{\bf k}$, $A_{\bf k}^2$, $B_{\bf k}^2$, $C_{\bf k}^2$, $X_{\bf k}Y_{\bf k}$ & $s$ ($d_{xy}$) \\
  $B_{1g}$ & 1 & -1 & -1 & 1 & 1 & -1 & -1 & 1 & line  & $A_{\bf k}B_{\bf k}$, $(X_{\bf k}+Y_{\bf k})Z_{\bf k}$ & $d_{(x+y)z}$  \\
  $B_{2g}$ & 1 & -1 & 1 & -1 & 1 & -1 & 1 & -1 & line & $A_{\bf k}C_{\bf k}$, $X_{\bf k}^2-Y_{\bf k}^2$  & $d_{x^2-y^2}$  \\
  $B_{3g}$ & 1 & 1 & -1 & -1 & 1 & 1 & -1 & -1 & line & $B_{\bf k}C_{\bf k}$, $(X_{\bf k}-Y_{\bf k})Z_{\bf k}$ & $d_{(x-y)z}$ \\
  $A_{1u}$ & 1 & 1 & 1 & 1 & -1 & -1 & -1 & -1 &  none & $A_{\bf k}B_{\bf k}C_{\bf k}$, $(X_{\bf k}^2-Y_{\bf k}^2)Z_{\bf k}$ & $f$ \\
  $B_{1u}$ & 1 & -1 & -1 & 1 & -1 & 1 & 1 & -1 & line  & $C_{\bf k}$, $X_{\bf k}+Y_{\bf k}$ & $p_{(x-y)}$  \\
  $B_{2u}$ & 1 & -1 & 1 & -1 & -1 & 1 & -1 & 1 & line  & $B_{\bf k}$, $Z_{\bf k}$ & $p_{z}$  \\
  $B_{3u}$ & 1 & 1 & -1 & -1 & -1 & -1 & 1 & 1 & line  & $A_{\bf k}$, $X_{\bf k}-Y_{\bf k}$ & $p_{(x+y)}$  \\
  \hline
\end{tabular}
\caption{The character table, symmetry required nodes in the
superconducting energy gap and some basis functions of the
irreducible representations of the point group $D_{2h}$. This is the
symmetry of the orthorhombic organic superconductors such as \Brn.
We argue in section \ref{sect:sc} that the experimental evidence
(see Figs. \ref{fig:Analytis} and \ref{fig:Carrington}) shows that
the superconducting order parameters of \Br and other orthorhombic
organic charge transfer salts transform like the $B_{2g}$
representation of $D_{2h}$. Thus the superconducting state may be
said to be `$d_{x^2-y^2}$'. In the colloquial names column we
include $d_{xy}$, parenthetically, as a name for the $A_{1g}$
representation. This is not intended to encourage the use of this
name but merely to stress that order parameters transforming like
the $B_{2g}$ representation of $D_{4h}$ (which have been invoked to
explain some experimental data; see, e.g. \cite{Izawa_ET}) transform
according to the $A_{1g}$ representation of $D_{2h}$ and are
therefore are not symmetry distinct from `$s$-wave' order
parameters. Note that in \Br the highly conducting plane is the
$a$-$c$ plane, and that the $x$ and $y$ axes are taken to lie along
the same directions as the $t$ hopping integrals (c.f., figure
\ref{fig:aniso}) therefore $k_a=k_x+k_y$ and $k_c=k_x-k_y$. The
functions $1_{\bf k}$, $X_{\bf k}$, $Y_{\bf k}$, $Z_{\bf k}$,
$A_{\bf k}$, $B_{\bf k}$ and $C_{\bf k}$ may be any functions which
transform, respectively, as 1, $k_x$, $k_y$, $k_z$, $k_a$, $k_b$ and
$k_c$ under the operations of the group and satisfy translational
symmetry. The operations of the group are the identity ($E$),
rotation by $\pi$ about the $a$, $b$ and $c$ axes (respectively
$C_2^a$, $C_2^b$ and $C_2^a$), inversion ($i$) and reflection
through the $ab$, $ac$ and $bc$ planes (respectively $\sigma^{ab}$,
$\sigma^{ac}$ and $\sigma^{bc}$). The character is the trace of any
matrix that can represent the operation in that irreducible
representation (see \cite{Tinkham-group} for a detailed discussion).
However, for the one-dimensional representations relevant here the
character has more physical interpretation: the character is the
sign introduced in the order parameter (and wavefunction) by that
operation. For example, any order parameter transforming according
to the $A_{1g}$ representation will by unchanged by any of the
operations of the group. On the other hand if the order parameter
transforms according to the $B_{2g}$ representation (as does, for
example, $\Delta_k\sim\cos k_x-\cos k_y$) then the order parameter
changes sign under rotation by $\pi$ about the $a$ ($x+y$) or $c$
($x-y$) axes and reflection through the $bc$ ($x=y$) and $ab$
($x=-y$) planes.}\label{tab:D2h}
\end{table}

\subsubsection{Spatial symmetry (`$s$-wave' versus `$d$-wave')}\label{sect:sym}

The organic charge transfer salts form orthorhombic, monoclinic and
triclinic crystals. This can lead to important differences between
their superconducting states \cite{group,q1d-phenom}. Orthorhombic
crystals such as \Br (in which the highly conducting plane is the
$a$-$c$ plane) are described by the $D_{2h}$ point group. There are
four irreducible representations of $D_{2h}$ which correspond to
singlet superconductivity (see table \ref{tab:D2h}). Given the
layered structure of the crystals superconductivity transforming as
either the $B_{1g}$ or $B_{3g}$ irreducible representations is
unlikely \cite{group,James_adv_phys}. Therefore, our task is to
differentiate between superconducting order parameters transforming
like the $A_{1g}$ representation (which is often referred to as
`$s$-wave' superconductivity) and superconducting states that
transform according to the $B_{2g}$ representation (or
`$d_{x^2-y^2}$' superconductivity).

There are even less symmetry distinct superconducting states in
monoclinic crystals such as \NCS (for which the highly conducting
plane is the $b$-$c$ plane) which are described by the $C_{2h}$
point group. There are two irreducible representations of $C_{2h}$
which correspond to singlet superconductivity (see table
\ref{tab:C2h}). Therefore, the singlet order parameter must
transform like either the $A_{g}$ representation (which we may refer
to as `$s$-wave' superconductivity) or the $B_{g}$ representation
(which we may refer to as `$d$-wave'
superconductivity\footnote{Although the only symmetry required node
for states transforming according to the $B_g$ representation lies
along the $c$-axis \cite{group} and so such states will not, in
general, be the same as the `$d_{x^2-y^2}$' state described above}).

\begin{table}[t]
\begin{tabular}{|l|cccc|c|c|c|c|}
  \hline
  Irrep & $E$ & $C_2^c$ & $i$ & $\sigma^{ab}$ & \begin{tabular}{c}
    Required \\
    nodes
  \end{tabular}
& Example basis functions
  & 
    Colloquial 
    names
  \\
  \hline
  $A_{g}$ & 1 & 1 & 1 & 1 & none & $1_{\bf k}$, $A_{\bf k}^2$, $B_{\bf k}^2$, $C_{\bf k}^2$, $X_{\bf k}Y_{\bf k}$ & $s$ ($d_{xy}$) \\
  $B_{g}$ & 1 & -1 & 1 & -1 & line & $A_{\bf k}C_{\bf k}$, $(X_{\bf k}-Y_{\bf k})Z_{\bf k}$, $B_{\bf k}C_{\bf k}$, $X_{\bf k}^2-Y_{\bf k}^2$  & $d$  \\
  $A_{u}$ & 1 & 1 & -1 & -1 & none & $A_{\bf k}$, $B_{\bf k}$, $X_{\bf k}+Y_{\bf k}$ & $p_{x+y}$ \\
  $B_{u}$ & 1 & -1 & -1 & 1 & line & $C_{\bf k}$, $X_{\bf k}-Y_{\bf k}$ & $p_{x-y}$  \\
  \hline
\end{tabular}
\caption{The character table, symmetry required nodes and some basis
functions of the even parity irreducible representations of the
point group $C_{2h}$. This represents the symmetry of the monoclinic
organic superconductors such as \NCSn. We argue in section
\ref{sect:sc} that the experimental evidence (see Figs.
\ref{fig:Analytis} and \ref{fig:Carrington}) shows that the
superconducting order parameters of \NCS and other monoclinic
organic charge transfer salts transform like the $B_{g}$
representation of $C_{2h}$. Thus the superconducting state may be
said to be `$d$-wave'. In the colloquial names column we include
$d_{xy}$, parenthetically, as a name for the $A_{1g}$
representation. This is not intended to encourage the use of this
name but merely to stress that order parameters transforming like
the $B_{2g}$ representation of $D_{4h}$ (which have been invoked to
explain some experimental data; see, e.g. \cite{Izawa_ET}) transform
according to the $A_{1g}$ representation of $C_{2h}$ are therefore
are not symmetry distinct from `$s$-wave' order parameters. Note
that in \NCS the highly conducting plane is the $b$-$c$ plane, and
that the $x$ and $y$ axes are taken to lie along the same directions
as the $t$ hopping integrals (c.f., figure \ref{fig:aniso})
therefore $k_b=k_x+k_y$ and $k_c=k_x-k_y$. The functions $1_{\bf
k}$, $X_{\bf k}$, $Y_{\bf k}$, $Z_{\bf k}$, $A_{\bf k}$, $B_{\bf k}$
and $C_{\bf k}$ may be any functions which transform, respectively,
as 1, $k_x$, $k_y$, $k_z$, $k_a$, $k_b$ and $k_c$ under the
operations of the group and satisfy translational symmetry. A brief
explanation of characters is given in the caption to table
\ref{tab:D2h}.}\label{tab:C2h}
\end{table}

One way to differentiate between `$s$-wave' and `$d$-wave'
states\footnote{From hereon we use the phrase `$d$-wave' to refer to
both the $B_g$ representation of $C_{2h}$ and the $B_{2g}$
representation of $D_{2h}$ unless we explicitly state otherwise.} is
to measure the \emph{low temperature} behaviour of thermodynamic
variables. For an `$s$-wave' state the superconducting gap is finite
at every point on the Fermi surface. As excitations may only take
place at energies within  about $k_BT$ of the Fermi energy
thermodynamic properties are exponentially activated at low
temperatures. For example, the heat capacity
$C_v\propto\exp(-\Delta_0/k_BT)$ for $T\ll T_c$, where $\Delta_0$ is
the minimum value of the magnitude of the superconducting gap at the
Fermi surface at $T=0$. In contrast for a `$d_{x^2-y^2}$' state
symmetry requires that the superconducting order parameter vanishes
along (at) four lines (points) on a three (two) dimensional Fermi
surface. These lines (points) are known as nodes. At the nodes there
are excitations with arbitrarily low energies. Thus although density
of states, $D(\epsilon)$, does vanish at the Fermi energy,
$\epsilon_F$, in a `$d$-wave' superconductor, the DOS grows linearly
as we move away from the Fermi energy, $D(\epsilon)\propto
|\epsilon-\epsilon_F|$ for $|\epsilon-\epsilon_F|\ll|\Delta|$. As,
for example, $C_v/T\sim D(\epsilon)$ it follows that at \emph{low
temperatures} $C_v\propto T^2$ in a `$d$-wave' superconductor.

Measurements of the temperature dependence of the heat capacity
\cite{heat-capacity}, penetration depth \cite{Carrington} and
nuclear spin lattice relaxation rate
\cite{de_Soto,Mayaffre,NMRreview} have produced apparently rather
contradictory results; with some groups arguing that their results
provide evidence for fully gapped `$s$-wave' pairing and other
groups arguing that their data favours states with nodes such as
`$d$-wave' superconductivity. (We recently reviewed this data in
rather more detail in \cite{disorder}.) However, it is important to
note that many of these experiments were not performed in the $T\ll
T_c$ limit and therefore do not strongly differentiate between
exponentially activated and power law behaviours. The only
experiment, which we are aware of, reporting data below about 20\%
of $T_c$ is the penetration depth, $\lambda$, study of Carrington
\etal \cite{Carrington} which reports data down to about 1\% of
$T_c$. They found a power law behaviour, but with a peculiar
$\lambda(T)-\lambda(0)\propto T^{3/2}$ behaviour\footnote{One
expects $\lambda(T)-\lambda(0)\propto T$ for line (point) nodes and
a three- (two-) dimensional Fermi surface and
$\lambda(T)-\lambda(0)\propto T^2$ for point nodes on a
three-dimensional Fermi surface \cite{Sigrist&Ueda}.} (see figure
\ref{fig:Carrington}). The correct interpretation of this
temperature dependence is not clear at the current time, except to
note that their data is quite inconsistent with an `$s$-wave' order
parameter.

\begin{figure}[t]
\centering
\includegraphics*[width=.7\textwidth]{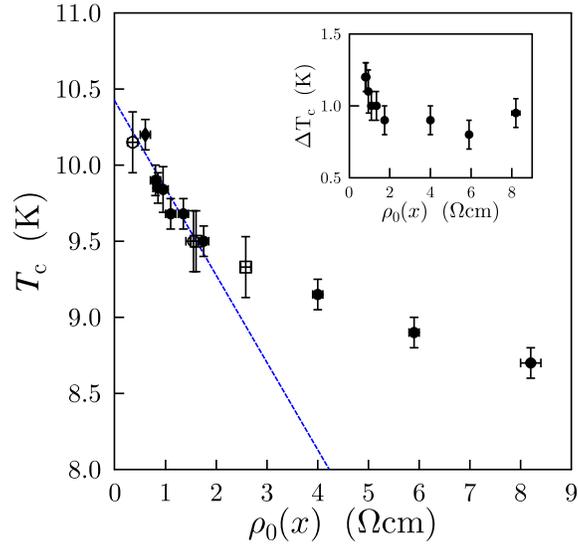}
\caption[]{Experimental evidence for unconventional
superconductivity in half-filled quasi-two-dimensional organic
charge transfer salts from measurements of suppression of $T_c$ by
disorder (taken from \cite{Analytis}). It shows the variation of the
critical temperature, $T_c$, and residual resistivity, $\rho_0$, of
\NCS as disorder is introduced by irradiating the sample. The line
shows the prediction of the Abrikosov-Gorkov theory (\ref{eqn:AG})
for a `non-$s$-wave' superconducting order parameter (that
transforms as a non-trivial representation). In contrast for an
`$s$-wave' order parameter Anderson's theorem \cite{AndersonTheorem}
predicts that the critical temperature is not suppressed by low
levels of disorder. Thus, the strong initial suppression of $T_c$ as
$\rho_0$ (which is inversely proportional to the electron-impurity
scattering rate) increases is strongly suggestive of a
`non-$s$-wave' order parameter. However, it is not currently
understood why $T_c$ is greater than is predicted by the
Abrikosov-Gorkov theory for $\rho_0\gtrsim2~\Omega$cm.}
\label{fig:Analytis}
\end{figure}

\begin{figure}[t]
\centering
\includegraphics*[width=.7\textwidth]{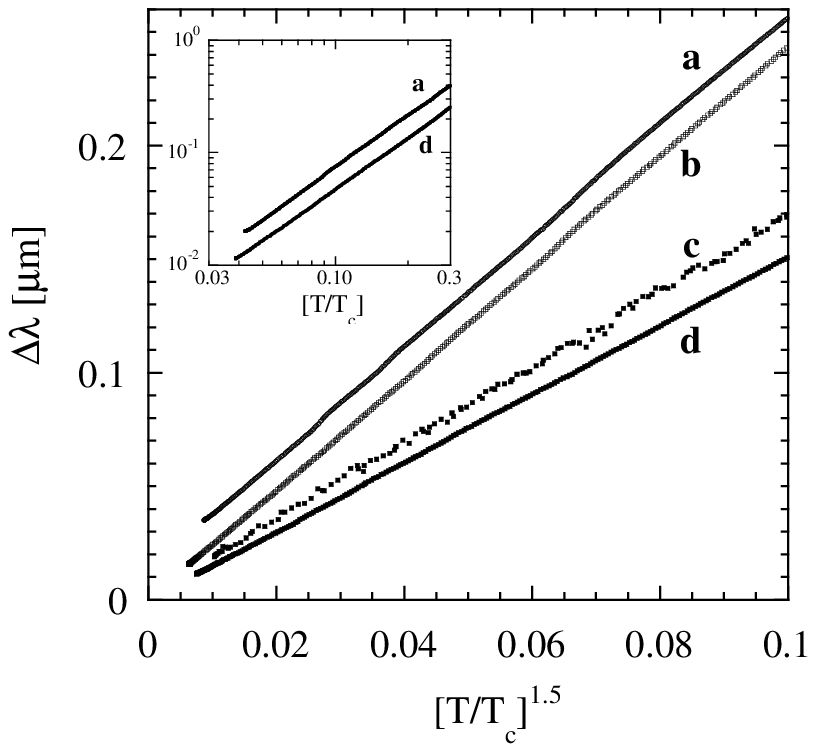}
\caption[]{Experimental evidence for unconventional
superconductivity in half-filled quasi-two-dimensional organic
charge transfer salts from the temperature dependence of the
penetration depth (relative to the penetration depth extrapolated to
$T=0$). The data is taken form \cite{Carrington}. Samples a and b
are \Brn, while samples c and d are \NCS (note that the non-linear
scale on the abscissa and that the data have been offset for
clarity). The inset shows the data for samples a and d on a log
scale. It is clear that $\Delta\lambda\sim(T/T_c)^\frac{3}{2}$. For
an `$s$-wave' gap (transforming like the identity representation)
one expects $\Delta\lambda\sim\exp(-k_BT/\Delta_0)$ at low
temperatures whereas one expects a power law behaviour when there
are nodes in the gap. However, there remains an unexplained feature
in this data: the power law is not what is expected in simple
theories where one expects $\Delta\lambda$ to vary either linearly
or quadratically with temperature at low temperatures. Experiments
such as this and that in figure \ref{fig:Analytis} suggest that the
pairing state of the organic charge transfer salts has a lower
symmetry than the crystal. In particular we argue in section
\ref{sect:sc} that orthorhombic materials, such as \Brn, have order
parameters which transform like the $B_{2g}$ representation of
$D_{2h}$ and so may be termed `$d_{x^2-y^2}$' superconductors. We
also argue that monoclinic materials, such as \NCSn, have order
parameters which transform like the $B_{g}$ representation of
$C_{2h}$ and so they may be termed `$d$-wave'
superconductors.}\label{fig:Carrington}
\end{figure}

In the cuprates phase sensitive probes, especially tunnelling
\cite{VanHarlingen} and scanning tunnelling microscopy (STM)
\cite{Davis} experiments have been particularly important for
determining the pairing symmetry. However, such experiments have not
been reliably performed in the organic charge transfer salts (see
\cite{disorder} for a critical review of the experiments that have
been performed to date.)

An important distinction between `$s$-wave' superconductors and
`non-$s$-wave' superconductors is the effect of non-magnetic
disorder. When a quasiparticle is scattered by  a non-magnetic
impurity its momentum is changed by a random amount. However, Fermi
statistics dictate that quasiparticles may only scatter from and to
states near the Fermi surface. Therefore, in a superconductor
impurity scattering has the effect of averaging the order parameter
over the Fermi surface \cite{disorder,Mineev&Samokhin}. For states
transforming as the trivial representation (e.g., the $A_{1g}$
representation of $D_{2h}$, see table \ref{tab:D2h}) the average of
the order parameter over the Fermi surface will, in general, be
non-zero, and hence $T_c$ is not suppressed \cite{AndersonTheorem}.
However, for order parameters which transform under the operations
of the point group like any representation other than the trivial
representation, that is for all `non-$s$-wave' states, the average
of the order parameter over the Fermi surface must vanish by
symmetry. Therefore, it can be shown \cite{Mineev&Samokhin,Larkin}
that non-magnetic disorder suppresses the critical temperature of
unconventional superconductors according to the Abrikosov-Gorkov
\cite{AG} formula,
\begin{eqnarray}
\ln \left(\frac{T_{c0}}{T_{c}} \right) = \psi\left( \frac{1}{2} +
\frac{\hbar}{4\pi k_BT_{c}}\frac{1}{\tau} \right) - \psi\left(
\frac{1}{2} \right), \label{eqn:AG}
\end{eqnarray}
where $T_{c0}$ is the superconducting critical temperature in the
pure system, $\psi(x)$ is the digamma function and $\tau$ is the
quasiparticle lifetime due to scattering from impurities. It was
recently pointed out that a large number of experiments in the
literature show that in the organic charge transfer salts presumably
non-magnetic impurities suppress $T_c$ in just the way predicted by
the Abrikosov-Gorkov formula \cite{disorder}. This has led to new
experiments specifically designed to investigate the role of
disorder in organic charge transfer salts \cite{Analytis}. These
experiments have, however, not conclusively resolved the pairing
symmetry. Indeed these experiments have resulted in a new puzzle.
Analytis \etal \cite{Analytis} irradiated samples of \NCS with both
protons and x-rays (both leading to similar results). Fig
\ref{fig:Analytis} shows the observed critical temperature plotted
against the residual resistivity of the sample. While this initially
follows the Abrikosov-Gorkov curve, for larger irradiation doses,
the suppression of $T_c$ is weaker than predicted. While this
behaviour seems rather inconsistent with a pure `$s$-wave' order
parameter a detailed explanation of this phenomena is  lacking.

The vortex lattice in the Abrikosov phase can yield important
information about the pairing symmetry. In particular for an
isotropic order parameter one expects a triangular vortex lattice
\cite{Abrikosov-nobel,Annett}. While, for an anisotropic order
parameter either a square or a triangular lattice may occur as the
energy difference between the square and triangular vortex lattices
is less than 1\% even for an isotropic order parameter
\cite{Annett,Abrikosov-nobel}. In particular square vortex lattices
are found in \SROn, UPt$_3$ and \YBCO \cite{Annett}, all of which
are believed to be unconventional superconductors. In the organic
charge transfer salts a triangular lattice is observed
\cite{PrattICSM04} by muon spin relaxation ($\mu$SR) experiments,
which does not give a strong indication of what the pairing symmetry
is.

We believe that a `$d_{x^2-y^2}$' order parameter transforming as
the $B_{2g}$ representation of $D_{2h}$ in the orthorhombic
materials and a `$d$-wave' order parameter transforming as the $B_g$
representation of $C_{2h}$ in the monoclinic crystals \cite{group}
is most consistent with currently available data. However, there is
no clear `smoking gun' experiment supporting this conclusion. We
eagerly await such an experiment and in the next section we review
some of the further experiments that could be used to probe the
pairing symmetry.

\subsection{Possible probes of the pairing
symmetry}\label{sect:probes}

We will now briefly discuss potential experiments that might provide
an unambiguous determination of the pairing symmetry. Particular
attention should be paid to directional probes such as ultrasonic
attenuation, thermal conductivity and experiments on single crystals
in a magnetic field. Directional probes have yielded important
information about the gap structure in the cuprates
\cite{VanHarlingen}, UPt$_3$ \cite{Joynt&Tallefer} and \SRO
\cite{Meano&Mackenzie}, so we will review what has been learnt by
such methods these materials.

\subsubsection{Effect of an orientated magnetic field in the conducting
plane on the heat capacity, nuclear spin lattice relaxation rate,
penetration depth and interlayer resistance}

In the vortex state of a superconductor with nodes in the energy gap
it can be shown that one effect of the applied magnetic field is to
introduce a Doppler shift to the quasiparticle energy due to
circulating supercurrents \cite{Volovik,Kubert}. Using this
semiclassical theory Vekhter \etal \cite{Vekhter} showed that  in a
`$d_{x^2-y^2}$' superconductor with a circular Fermi surface the
density of states at the Fermi level, $D(E_F,\alpha)$, where
$\alpha$ is the angle between the field and the antinodal direction,
has a four fold variation as a magnetic field is rotated around the
highly conducting plane. $D(E_F,\alpha)$ is maximal when ${\bf H}$
is aligned in the antinodal direction and minimal for ${\bf H}$
parallel to the nodes.

The variation in $D(E_F,\alpha)$ should be directly reflected in the
behaviour of the electronic contribution to the specific heat
$C_v(\alpha)$ which is predicted \cite{Vekhter} to show a four-fold
variation as ${\bf H}$ is rotated around the highly conducting
plane. Further Vekhter \etal predict that $C_v/T\propto\sqrt{H}T$
for the field in the nodal direction and that
$C_v/T\propto\sqrt{H}T^2$ for the field in the antinodal direction.
Similar effects are predicted for the ${\bf H}(\alpha)$ dependence
of the nuclear lattice relaxation rate $1/T_1T$, the penetration
depth $\lambda$ \cite{Vekhter} and the interlayer resistance
\cite{Bulaevskii99}.

Vekhter \etal also discuss the effect of lowering the symmetry of
the unit cell and thus introducing a small `$s$-wave' admixture to
the order parameter. They show that even a very small `$s$-wave'
component reduces the four-fold variation of $D(E_F,\alpha)$ to a
two-fold variation. However, experimentally great care would be
required to detect a two-fold variation in, for example, $C_v$ as
any slight mismanagement of the field out of the plane would cause a
similar two-fold variation because of the anisotropy between
$H_{c2\|}$ and $H_{c2\perp}$, where  $H_{c2\|}$ is the in plane
upper critical field and $H_{c2\perp}$ is the out of plane upper
critical field \cite{Deguchi}.

Deguchi \etal \cite{Deguchi} have observed four-fold variation in
$C_v(\alpha)$ in \SROn. However, they interpret this as evidence of
an anisotropic (but nodeless) gap on the $\gamma$ sheet of the Fermi
surface. In fact Deguchi \etal observed two different four-fold
observations as a function of the applied magnetic field strength.
For $|{\bf H}|\lesssim H_{c2\|}$ they observe an oscillation which
they attribute to the anisotropy of $H_{c2\|}$ in the basal plane of
\SROn. At intermediate fields there are no oscillations in
$C_v(\alpha)$ and for $H_{c1} < |{\bf H}| \ll H_{c2\|}$ Deguchi
\etal observe oscillations in $C_v(\alpha)$ that are $\pi/4$ out of
phase with those in the high field region. It is these low field
oscillations that appear to associated with the anisotropy of the
superconducting gap. In the organic charge transfer salts the in
plane upper critical field is Clogston--Chandrasekhar (or Pauli)
limited, \cite{Zuo} and there is therefore no anisotropy in
$H_{c2\|}(\alpha)$ so one should not expect the four-fold
oscillations in the high field region because of variations in
$H_{c2\|}(\alpha)$. However, this observation does indicate the ease
with which extraneous effects can enter this type of experiment and
therefore the care with which any such data needs to be interpreted.


Two related predictions are that for a gap with line nodes that
$C_v\sim H^\frac12T$ for $H\ne0$ and $T\ll T_c$ in a polycrystalline
sample \cite{Volovik,Won&Maki} and also that the heat capacity
scales \cite{Simon} as a function of $H^\frac12T$. Both of these
predictions have been confirmed for \YBCO \cite{YBCO_Cv} and
La$_{2-x}$Sr$_x$CuO$_4$ \cite{Fisher}.

None of the effects described in this section appear to have been
studied systematically in any of the organic charge transfer salts.
Clearly such experiments could be extremely powerful tools for
elucidating the structure of the gap in the organic charge transfer
salts.

\subsubsection{Thermal conductivity}

In the cuprates, UPt$_3$ and \SRO the thermal conductivity is
dominated by quasiparticles (rather than phonons) over a significant
temperature range below $T_c$. When quasiparticles dominate the
measurements the dependence of thermal conductivity of a type II
superconductor in the Abrikosov phase on the orientation of the
magnetic field can yield information about the structure of the
superconducting order parameter. For a superconductor with an
isotropic gap one expects that the thermal conductivity, $\kappa$
varies as $\cos\theta$ where $\theta$ is the angle between the heat
current and the magnetic field \cite{Maki}. This prediction has been
confirmed in  Nb \cite{Lowell}. With a thermal gradient along the
$a$- \cite{Yu} or $b$-axes \cite{Aubin} of \YBCO an additional four
fold variation in $\kappa(\theta)$ is observed when a magnetic field
is rotated in the basal plane. The thermal conductivity is maximal
at $\theta=\pi/4+n\pi/2$ for $n\in \mathbb{Z}$ and therefore
consistent with `$d_{x^2-y^2}$' superconductivity, i.e., the thermal
conductivity is maximal with the field aligned in the antinodal
direction \cite{Moreno}. In contrast in the (low temperature, low
field) B phase of UPt$_3$ only the two fold variation of
$\kappa(\theta)$ expected for an isotropic gap is observed when the
field is rotated in the basal plane \cite{Suderow}. This is
consistent with either an $E_{2u}$ or an $E_{1g}$ hybrid gap
structure \cite{Joynt&Tallefer}.

In \SRO a four fold anisotropy is observed in $\kappa(\theta)$ as
the field is rotated in the basal plane \cite{Izawa_SRO}. However,
this is believed to be a result of the tetragonal crystal structure
rather than a reflection of nodes (or anisotropies) in the gap
\cite{Meano&Mackenzie}. In particular in \SRO the fourfold
anisotropy is a much weaker than the effect in \YBCOn, and 20 times
small than that predicted \cite{Dahm} for a gap with vertical line
nodes.

Izawa \etal have measured the thermal conductivity of \NCS as a
magnetic field is rotated in the conducting ($b$-$c$) plane
\cite{Izawa_ET}. They observe a small (0.2\% for $T<0.6$~K) fourfold
variation in $\kappa(\theta)$. This fourfold anisotropy has its
maximum when the field is at 45$^\circ$ to the $b$ and $c$ axes
(i.e., along the $x$ and $y$ axes as defined in figure
\ref{fig:aniso}). Therefore, Izawa \etal propose that the
superconducting order parameter of \NCS has `$d_{xy}$' symmetry.
However, one should be cautious of this interpretation because a
`$d_{xy}$' order parameter transforms like the $A_g$ representation
of $C_{2h}$ (see table \ref{tab:C2h}). Therefore, the anisotropy
Izawa \etal observe is has the \emph{same} symmetry as the lattice.
Therefore, the observed anisotropy is compatible with variations
caused by the crystal lattice itself and the observed effect is
rather small. Indeed Izawa \etal also observed a large twofold
anisotropy, which they claim is ``mainly due to phonons". 

Furthermore, Izawa \etaln's interpretation is based on the
assumption that there is strong interlayer coupling, which may not
correct for \NCSn. In particular the theory used to interpret these
experiments requires that the superconductivity may be described by
a (highly anisotropic) three dimensional Ginzburg-Landau theory.
However, in \NCS there is a very real possibility that the layers
are Josephson coupled; in which case a three dimensional
Ginzburg-Landau theory could not be applied to these results. In
particular Izawa \etaln's interpretation of the data requires that
vortices are formed when the field is parallel to the highly
conducting layer.

The thermal conductivity of \Ga and \kFe has also been measured by
Tanatar and co-workers \cite{Tantar_BETS}. However, in this work the
field was not rotated in the conducting plane, so we will not
discuss it further.

\subsubsection{Ultrasonic attenuation}

Because the velocity of sound is much less than the velocity of an
electron at the Fermi surface ($v_s\ll v_F$), longitudinally
polarised ultrasound is only attenuated effectively by electrons
moving almost perpendicular to the direction of sound propagation
\cite{Tinkham}. This makes longitudinal ultrasound a powerful probe
of the anisotropy of the superconducting gap, indeed longitudinal
ultrasonic attenuation experiments were among the first to probe gap
anisotropy in conventional superconductors such as Sn \cite{Morse}.
Transverse ultrasound experiments are extremely sensitive to gap
anisotropies as the attenuation depends on both the direction of
sound propagation and the direction of the
polarisation.\footnote{For a clear explanation of this phenomena see
\cite{Meano&Mackenzie} and \cite{Ellman} and references therein.}
Such experiments have been important for determining the gap
structure of both UPt$_3$ \cite{Shivaram,Ellman} and \SRO
\cite{Talifer}.

Measurements of attenuation of longitudinal ultrasound perpendicular
to the highly conducting planes have been made in \NCS
\cite{ultraNCS,Simizu,Frikach}, \Br \cite{Frikach,ultraBr} and \Cl
\cite{Fournier}. A clear coupling to the electronic degrees of
freedom is seen as the crossover from the `bad-metal' to the Fermi
liquid is observed at temperatures corresponding to the maximum in
the resistivity. Indeed the electronic pressure-temperature phase
diagram of \Cl has been mapped out using ultrasound \cite{Fournier}.
However, none of these experiments were designed to measure the
anisotropy of the superconducting order parameter.

Ultrasound experiments are complicated by the small size and the
shape of single crystals of organic charge transfer salts
\cite{Simizu,Frikach}. However, for example, the crystal of \NCS
studied by Simizu \etal \cite{Simizu} (which was reported to be
$1.35\times5.06\times1.62$ mm$^3$) is rather similar size an shape
to the crystal of UPt$_3$ studied by Ellman \etal \cite{Ellman}
($1\times1\times2.7$ mm$^3$). Therefore, there does not seem to be
any intrinsic reason why transverse ultrasound could not be used to
study the order parameter of organic superconductors. Given the
importance of such experiments\cite{Shivaram,Ellman,Talifer,Morse}
in UPt$_3$, \SRO and Sn transverse ultrasound appear to be an
excellent tool to probe the structure of the gap in organic charge
transfer salts.

\subsection{Superfluid stiffness (evidence for a non-BCS groundstate and pairing
mechanism)}\label{sect:Pratt}

The defining property of a superconductor is the Meissner effect,
i.e., perfect diamagnetism \cite{Annett}. The strength of the
Meissner effect is measured by the superfluid stiffness, $D_s\equiv
c^2/4\pi\lambda^2$, where $\lambda$ is the penetration depth, as the
superfluid stiffness is the constant of proportionality between the
vector potential of an applied magnetic field and the induced
supercurrent. Thus the smaller the superfluid stiffness the `weaker'
the superconductivity. In the underdoped cuprates it is found that
$T_c \propto D_s$: this is the Uemura relation \cite{Uemura}.

A number of exotic mechanisms have been proposed to explain the
Uemura relation and it is natural in both the preformed pairs
scenario \cite{EK} and the resonating valence bond (RVB) theory
\cite{Zhang}. However, the Uemura relation in the underdoped
cuprates is even predicted within BCS theory \cite{Balazs}. In BCS
theory $D_s\propto n_s/m^*$, where $n_s$ is the density of electrons
in the superfluid condensate. However, we stress that there is no
way to directly measure $n_s$ and, in general, $n_s$ is not required
to be the same as the electron density, $n$, even at $T=0$ (for
example only about 10\% of the atoms form the condensate in $^4$He
at absolute zero \cite{Mook,Griffin}). However, BCS theory is a weak
coupling theory and $n_s=n$ at $T=0$. Recall that the doping, $x$,
is an implicit parameter when one compares $T_c$ and $D_s$ in the
underdoped cuprates. Therefore, the Uemura relation can be derived
from BCS theory by noting that $T_c\sim x$, which is observed
experimentally for small $x$, and that $D_s \propto n_s=n\propto x$
which assumes $m^*$ is independent of, or only weakly dependent on,
doping.

\begin{figure}[t]
\centering
\includegraphics*[width=.7\textwidth]{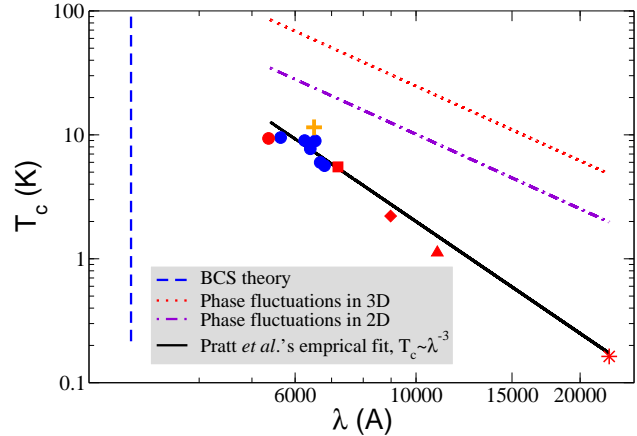}
\caption[]{The small superfluid stiffness observed in organic charge
transfer salts with low critical temperatures provides clear
evidence that the superconductivity is not described by BCS theory.
The predictions of the phase fluctuation theory proposed for the
cuprates by Emery and Kivelson \cite{EK} is also to overestimate the
critical temperature. This data is extremely surprising as the data
on the left corresponds to materials where the interactions are
stronger, e.g., the effect mass is larger \cite{constraints}. Thus
the superfluid stiffness $D_s\propto1/\lambda^2$, where $\lambda$ is
the zero temperature penetration depth, gets smaller (and hence less
like the BCS prediction) as we move away from the Mott transition.
This is unexpected as the spectral weight in the quasiparticle peak
increases as the correlations become weaker, so we might expect the
superfluid stiffness to be smallest near the Mott transition
\cite{Pratt&Blundell,RVB-organics}. Note that all of these materials
are half-filled \cite{constraints} so doping effects cannot be
invoked to explain the data. This figure is modified from
\cite{constraints}, while the data is from \textcolor{Red}{Pratt
\etal \cite{Pratt_nU}}, \textcolor{BurntOrange}{Lang \etal
\cite{Lang_penetration2}} and \textcolor{blue}{Larkin \etal
\cite{Larkin_penetration_depth}} (colour of citation matches colour
of data points). The figure shows data for
$\kappa$-ET$_2$Cu[N(CN)$_2$]Br ({\bf +}), $\kappa$-ET$_2$Cu(NCS)$_2$
at several pressures ($\bullet$), $\lambda$-BETS$_2$GaCl$_4$
($\blacksquare$), $\beta$-ET$_2$IBr$_2$ ($\blacklozenge$)
$\alpha$-ET$_2$NH$_4$Hg(NCS)$_4$ ($\blacktriangle$) and
$\kappa$-BETS$_2$GaCl$_4$ ($\bigstar$).%
} \label{fig:penetration}
\end{figure}

Pratt and coworkers \cite{Pratt_nU} have systematically measured the
critical temperatures and penetration depths of a large number of
organic charge transfer salts. This data, along with that of several
other groups \cite{Lang_penetration2,Larkin_penetration_depth} is
shown in figure \ref{fig:penetration}. It is important realise that
the implicit parameters in this plot are chemical substitution and
pressure, which do not change the filling factor: all of these
materials are half-filled \cite{constraints}. The role of pressure,
be it `chemical' or hydrostatic, in these materials is to drive them
away from the Mott transition \cite{Kanoda}. Generically, it is also
found that increasing the pressure lowers $T_c$. Therefore, the data
points on the left hand side of figure \ref{fig:penetration} are
closer to the Mott transition than those on the right hand side.

As there is no doping of the system BCS theory (and other weak
coupling theories) does not predict any change in $n_s$ as pressure
varies $T_c$ (whence the vertical line in figure
\ref{fig:penetration}). However, one might object that as we lower
pressure and drive towards the Mott transition $m^*$ increases. If
one includes this effect  the prediction is that the superfluid
stiffness is smallest closest to the Mott transition. This is the
\emph{opposite} behaviour to that seen experimentally. It seems
extremely unlikely that using Eliashberg theory to account for
strong coupling phononic effects will rectify this essential
disagreement with experiment \cite{constraints}. Therefore, Pratt
\etaln's data \cite{Pratt_nU} provides the clearest evidence (i) for
a non-phononic pairing mechanism and (ii) that weakly correlated
theories are insufficient to explain the observed superconductivity
in the organic charge transfer salts \cite{constraints}.

\section{Strongly correlated models of superconductivity: frustration and
RVB}\label{sect:SCtheory}

Most of the early work on superconductivity in the organic charge
transfer salts took weakly correlated approaches. A comprehensive
review of this work was recently published by Kuroki \cite{Kuroki},
and, rather than duplicating that effort here, we limit ourselves to
a few general comments. The two most studied weakly correlated
theories of superconductivity in the organic charge transfer salts
are the Eliashberg phononic pairing mechanism and the
spin-fluctuation pairing mechanism [with calculations most often
performed within the fluctuation-exchange (FLEX) approximation].
Neither of these approaches capture the Mott transition or the large
effective mass enhancement seen in the organic charge transfer
salts. Therefore, one must clearly go beyond a weakly correlated
description of the superconductivity in order to provide a complete
description of the full range of behaviours observed in the organic
charge transfer salts (c.f. sections \ref{sect:metal} and
\ref{sect:Mott}). Further, the small superfluid stiffness observed
in the low $T_c$ materials (figure \ref{fig:penetration} and section
\ref{sect:Pratt}) cannot be accounted for within weakly correlated
theories \cite{constraints}.

Recently a number of groups have proposed strongly correlated
theories of the organic charge transfer salts. These are based on
the Hubbard model described in section \ref{sect:hlos}. A number
different methods have been discussed
\cite{RVB-organics,unpublished,Liu,Zhang-organic,Younoki,Sahebsara,Kyung&Tremblay}
with rather similar results. Here we focus on the simplest of these
theories, the resonating valence bond (RVB) or gossamer
superconductor theory (the two names have been used interchangeably
in the literature).

The celebrated BCS wavefunction \cite{BCS} is
\begin{eqnarray}
|BCS\rangle = \prod_{\bf k} \left( u_{\bf k} + v_{\bf k}
\hat{c}_{{\bf k}\uparrow}^\dagger \hat{c}_{-{\bf
k}\downarrow}^\dagger \right) |0\rangle,
\end{eqnarray}
where $|0\rangle$ is the vacuum state and the $u_{\bf k}$ and the
$v_{\bf k}$ are variational parameters. The RVB wavefunction is a
projected BCS wavefunction,
\begin{eqnarray}
|RVB\rangle = \hat{P}_G|BCS\rangle,
\end{eqnarray}
where
\begin{eqnarray}
\hat{P}_G = \sum_i\left( 1-\alpha \hat{n}_{i\uparrow}
\hat{n}_{i\downarrow} \right)
\end{eqnarray}
is the partial Gutzwiller projector. For $\alpha=1$ the Gutzwiller
projector removes all double occupation from the wavefunction. The
$\alpha=1$ RVB state has been studied since the early days of
high-$T_c$ \cite{AndersonRVB,Zhang}. However, as no double
occupation can occur for $\alpha=1$ this state is always insulating
for half-filled systems and therefore will not give the correct
description of the organic charge transfer salts. However, if we
treat $\alpha$ as a variational parameter \cite{Laughlin,ZhangPRB}
then some double occupation is allowed, and superconducting and
metallic states may occur.

The simplest approach to this theory is to make the Gutzwiller
approximation (as well as the Hartree-Fock-Gorkov approximation
implicit in the BCS state) in which one enforces the constraints on
the fraction of doubly occupied sites only on average. This allows
one to derive a strongly correlated mean-field theory with only a
few parameters \cite{Laughlin,ZhangPRB}. Studies of this theory
\cite{unpublished,Zhang-organic,RVB-organics} have shown that the
superconducting state: (i) undergoes a first-order Mott transition
as is seen experimentally (in the mean-field theory this is of the
Brinkman-Rice type \cite{Brinkman-Rice}); (ii) has quasiparticles
with a large effective masses which are strongly enhanced near the
Mott transition as is seen in experiments; and (iii) has a strongly
suppressed superfluid stiffness. The RVB theory also predicts that
there is a pseudogap above the superconducting state in agreement
with NMR experiments (see section \ref{sect:pseudoTheory}).
Furthermore, it has already been shown that the insulating state of
the RVB theory supports both N\'eel ordered states
\cite{Zhang-organic,Sahebsara,Kyung&Tremblay} like that observed in
the insulating state of \Cl and spin-liquid states
\cite{RVB-organics,Liu,Kyung&Tremblay} like those observed in the
insulating state of \CN and \dmit (see section \ref{sect:spinL}).
However, to date there as been relatively little work
\cite{Kyung&Tremblay} on the comparative stability of these states
as the frustration is varied. Another important task for these
theories yet to be reported is to give a detailed explanation of the
small superfluid stiffness seen in the low-$T_c$ materials (see
figure \ref{fig:penetration} and section \ref{sect:Pratt}).

\section{The phase diagram of the Hubbard model on the anisotropic triangular
lattice}\label{sect:pd-tri}

Eight years ago \cite{Ross-review} one of us proposed a speculative
zero temperature phase diagram for the Hubbard model on an
anisotropic triangular lattice. We now know significantly more about
the phase diagram because there have been many studies of this model
utilising many different approximation schemes and numerical
techniques. In figure \ref{fig:tri-lat-pd} we sketch a schematic
phase diagram which summarises current knowledge.

\subsection{The Mott transition}

\begin{figure}
\centering
\includegraphics*[width=.7\textwidth]{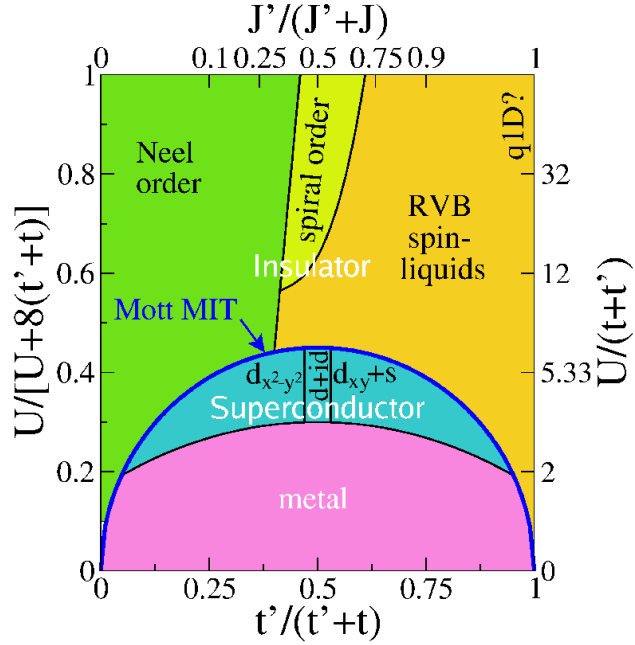}
\caption[]{Schematic diagram of the proposed zero temperature phase
diagram of the Hubbard model on the anisotropic triangular lattice
at half filling. We argue throughout this review that this model
provides a qualitative description of the $\beta$, $\beta'$,
$\kappa$ and $\lambda$ phase organic charge transfer salts. The
central feature of the phase diagram is the Mott transition. We have
argued that superconductivity occurs on the metallic side of the
Mott transition. It has been proposed that the symmetry of the
superconducting state varies as the frustration is varied and that
the superconducting state breaks time reversal symmetry when the
lattice is approximately hexagonal ($t\sim t'$)
\cite{group,unpublished}. We label the superconducting states by
their symmetries (the colloquial names are given a more formal basis
in table \ref{tab:C2v}). It should be stressed that the proposed
superconducting states are strongly correlated and RVB like rather
than weakly correlated BCS states
\cite{Kyung&Tremblay,Sahebsara,Liu,Zhang-organic,RVB-organics,unpublished}.
As the correlations, which mediate the superconductivity, are
reduced we recover a metallic groundstate. For $U\gg t,t'$ the
Hubbard model at half filling is insulating and the spin degrees of
freedom can be described by the Heisenberg model. On the basis of
the calculations reported in
\cite{series-expanisions,Jaime-spin-wave,Cheung,chubukov,120} we
have also argued that the insulating states should change from
N\'eel order to spiral ordered states to a spin-liquid as $t'/t$ is
increased, i.e., as the frustration is varied. Near the Mott
transition corrections to the Heisenberg model, such as ring
exchange are important. This may stabilise the spin-liquid
\cite{MotrunichPRB05}. This phase diagram includes all of the phases
observed in the layered organic charge transfer salts at low
temperatures. We have shown elsewhere in this review (particularly
sections \ref{sect:metal}, \ref{sect:Mott} and \ref{sect:SCtheory})
that the finite temperature behaviour of the half-filled layered
organic charge transfer salts is also described by the Hubbard model
on an anisotropic triangular lattice. Hence we conclude that that
the Hubbard model on the anisotropic triangular lattice provides an
excellent qualitative description of these materials.}
\label{fig:tri-lat-pd}
\end{figure}

Only a few points on the phase diagram are known exactly. Because of
perfect nesting the Mott metal-insulator transition occurs at an
infinitesimal $U$ for both the square lattice ($t'=0$) and isolated
1D chains ($t=0$) \cite{Lieb&Wu}. For the isotropic triangular (or
more correctly, hexagonal) lattice ($t'=t$) there is no nesting at
half filling and so the Mott transition occurs at a finite $U$. The
critical $U$ has been estimated by a variety of methods
\cite{RVB-organics,NaxCoO,Picket-DMFT,exact} and seems to be about
$U=10t-15t$. We therefore sketch the Mott transition as passing
smoothly between these three points in our phase diagram. Several
approximations suggest that the Mott transition is first order
throughout the (zero temperature) phase diagram, however the recent
work of Imada \etal \cite{Imada,Misawa} and the critical exponents
measured by Kawagwa \etal \cite{Kagawa} has called this into
question. In particular Imada \etal propose that for some
frustrations the transition is first order while for others there is
a (marginal) quantum critical point. All of the other phase
transitions shown in the schematic phase diagram (figure
\ref{fig:tri-lat-pd}) are thought to the second order.

\begin{figure}
\centering
\includegraphics*[width=.7\textwidth]{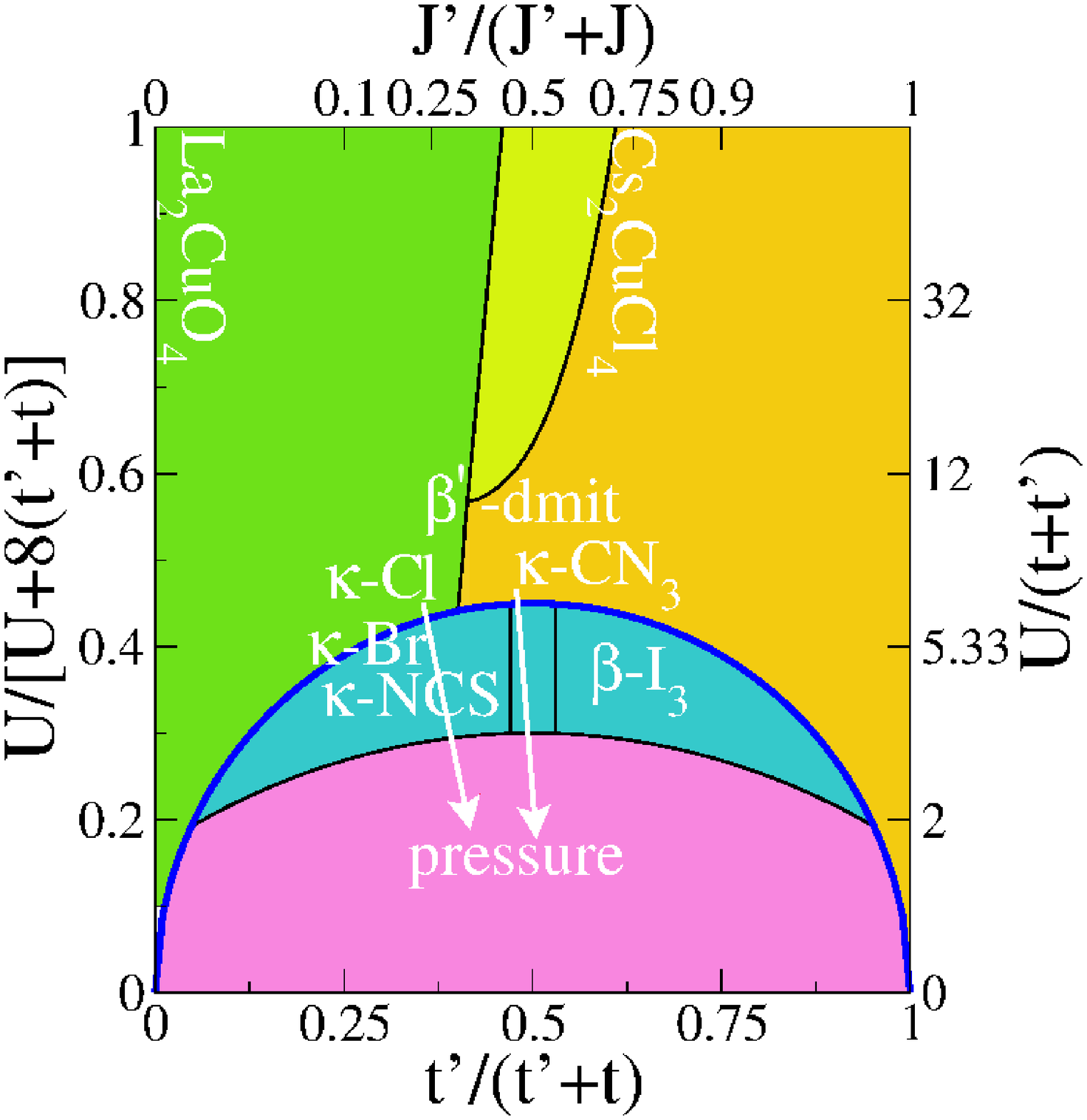}
\caption[]{The same phase diagram as is shown in figure
\ref{fig:tri-lat-pd} with the labels removed (for clarity) and the
names of some materials added. The position of name indicates our
guestimate of where that material sits on the proposed phase
diagram, the white arrows indicate the proposed effect of pressure.
This is very rough and accurately determining the details of the
parameter values corresponding to specific materials at particular
pressures is a major challenge (see in particular section
\ref{sect:cal-params}). The following shorthand is used in the
figure: $\kappa$-Cl is \Cln, $\kappa$-CN$_3$ is \CNn, $\kappa$-Br is
\Brn, $\kappa$-NCS is \NCSn, $\beta$-I$_3$ is \bI and $\beta$'-dmit
is \dmitn. La$_2$CuO$_2$, the parent compound of the cuprates, is
shown with a large $U$ and on the square lattice ($t'=0$) to
emphasise the similarities between the organics and the cuprates
\cite{Ross-science}. It is known experimentally that $J'/J\approx3$
in Cs$_2$CuCl$_4$ \cite{CCCmeasureJ}. Cs$_2$CuCl$_4$ has a spirally
ordered ground state at very low temperatures, but  shows a region
of spin-liquid like behaviour at higher temperatures
\cite{CCCspin-liquid} which may indicate that the spin-liquid state
is energetically close to the ground state and that coupling in the
third dimension plays an important role in the real material. \CN
shows a spin-liquid ground state \cite{Shimizu}, while the \dmit
salts seem to be on the boarder between magnetic ordering and
spin-liquid behaviour. The nature of this magnetic ordering is not
yet clear. Pressure drives \Cl from a the N\'eel state via the
superconducting state to normal metal. While \Br is so close to the
Mott transition that it can be driven insulating by deuteration of
the ET molecule. Huckel calculations suggest that \bI (and other
$\beta$ phase salts) have $t'>t$ \cite{Huckel-beta}. If our
speculative phase diagram is correct then this suggests that they
have a different pairing symmetry to the $\kappa$-phase materials
\cite{unpublished,group}. This also suggests that the
superconducting states of \CN and \dmit may break time reversal
symmetry \cite{unpublished,group}.} \label{fig:tri-lat-pd-mat}
\end{figure}

\subsection{The superconducting states}

A number of studies based on strongly correlated theories
\cite{RVB-organics,Liu,Zhang-organic,Sahebsara,unpublished,Kyung&Tremblay}
suggest that superconductivity is realised on the metallic side of
the Mott transition. This superconductivity is mediated by
antiferromagnetic interactions which occur because of superexchange.
As the superexchange interactions are only relevant in the large $U$
limit, for small $U$ this interaction will vanish and so will the
superconductivity, leaving a metallic state for $W\gg U$.

The symmetry of the anisotropic triangular lattice is represented by
the $C_{2v}$ point group, which is summarised in table
\ref{tab:C2v}. Various calculation methods indicate that in the
small $t'/t$ limit the superconducting order parameter transforms
like the $A_2$ representation of $C_{2v}$ \cite{group,unpublished}.
This is colloquially referred to as `$d_{x^2-y^2}$'
superconductivity (see section \ref{sect:symetry} for a discussion
of this nomenclature). FLEX \cite{Kuroki} and weak-coupling
renormalisation-group \cite{Marston} studies also predict that the
superconducting state has this symmetry. However, for $t=t'$ the
model has $C_{6v}$ symmetry. Here `$d_{x^2-y^2}$' state belongs to
the two-dimensional $E_{2}$ representation. This has recently led to
the proposal that for $t\sim t'$ a `$d+id$' state (which is a mixed
state transforming like a complex combination of both the $A_1$ and
$A_2$ representations of $C_{2v}$) is realised
\cite{group,unpublished}. These arguments also suggest that in the
large $t'/t$ limit the state transforming like the $A_1$
representation of $C_{2v}$, which may be colloquially referred to as
a `$d_{xy}+s$' state, occurs \cite{group,unpublished}. We therefore
sketch these different superconducting states in figure
\ref{fig:tri-lat-pd}.

\begin{table}[t]
\begin{tabular}{|l|cccc|c|c|c|c|}
  \hline
  Irrep &  $E$ & $C_2$ & $\sigma^+$ & $\sigma^-$ & Required nodes
& Example basis functions
  & Colloquial names \\
  \hline
  $A_{1}$ & 1 & 1 & 1 & 1 & none & $1_{\bf k}$, $X_{\bf k}Y_{\bf k}$ & $s$, $d_{xy}$, $s+d_{xy}$ \\
  $A_{2}$ & 1 & 1 & -1 & -1 & line  & $X_{\bf k}^2-Y_{\bf k}^2$ & $d_{x^2-y^2}$  \\
  $B_{1}$ & 1 & -1 & 1 & -1 & line & $X_{\bf k}+Y_{\bf k}$  & $p_{x+y}$  \\
  $B_{2}$ & 1 & -1 & -1 & 1 & line & $X_{\bf k}-Y_{\bf k}$ & $p_{x-y}$ \\
  \hline
\end{tabular}
\caption{The character table, symmetry required nodes and some basis
functions of the irreducible representations of the point group
$C_{2v}$. $C_{2v}$ represents the symmetry of the anisotropic
triangular lattice. The functions $1_{\bf k}$, $X_{\bf k}$ and
$Y_{\bf k}$ may be any functions which transform, respectively, as
1, $k_x$ and $k_y$ under the operations of the group and satisfy
translational symmetry. The operations in the group are the
identity, $E$, rotation about the $z$-axis by $\pi$, $C_2$, and
reflection through the planes $k_x=\pm k_y$, $\sigma^\pm$.  A brief
explanation of characters is given in the caption to table
\ref{tab:D2h}. Note that inversion symmetry is not an operation of
the group. This might suggest that singlet and triplet states are
not differentiated by symmetry. However, as we are discussing the
symmetry of a two dimensional lattice, rotation by $\pi$ (which is
an operation in the group) differentiates between singlet and
triplet states.}\label{tab:C2v}
\end{table}

\subsection{The insulating states}

We now turn our attention to the insulating state. We begin by
considering the large $U/W$ limit, in which our Hubbard model
reduceds to the Heisenberg model on the anisotropic triangular
lattice. This model contains two parameters, $J=4t^2/U$ and
$J'=4t'^2/U$ which occur because of of superexchange: a second order
perturbation effect. All higher order process can be neglected
because we are in the large $U$ limit. For small $J'/J$ it is well
established that the N\'eel state is realised. Series expansions
\cite{series-expanisions} show that the N\'eel state is stable for
$J'/J<0.7$. For $J=0$ we have uncoupled one-dimensional chains along
the diagonal on the unit cell. The Heisenberg model can be solved
exactly in one-dimension and the ground state is a spin-liquid with
deconfined spinon excitations \cite{Tsvelik}. The question of what
happens when there is a finite coupling to a second dimension
(finite $J$) is extremely subtle. However, recent work
\cite{Estler-Balents,series-expanisions,Younoki} suggests that this
physics does survive to a significant degree in frustrated systems
and that a spin-liquid occurs in some parts of the phase diagram
with $J'>J$. Therefore, we label the large $J'/J$ region of the
phase diagram `q1D?'.

Various theoretical studies of the Heisenberg model on the
anisotropic triangular lattice have been performed. The methods used
include linear spin wave theory \cite{Jaime-spin-wave} and large-$N$
mean field theory \cite{Cheung}. Both of these methods indicate that
when frustration destroys N\'eel order [which has the ordering
wavevector $\vec q=(\pi,\pi)$] a spiral state [with ordering
wavevector $\vec q=(q,q)$] occurs. In the spiral state $q$ varies
from $\pi$, at the critical frustration where the N\'eel state gives
way to spiral order, to $\pi/2$ at large frustration which is the
classical value for uncoupled chains. At $J'=J$ these theories give
$q=3\pi/2$ which describes the `classical $120^\circ$-state': the
classical solution of the isotropic triangular lattice
\cite{chubukov}. In the classical $120^\circ$-state the spins on
neighbouring sites align at an angle $120^\circ$ apart from their
nearest neighbours. It is widely believed that the quantum analogue
of `$120^\circ$-state' is the true ground state of the Heisenberg
model on the isotropic triangular lattice
\cite{series-expanisions,120}. By analogy it may be argued that
quantum-spiral states are the ground states of the Heisenberg model
on the anisotropic triangular lattice in the regime $J'\sim J$. This
is certainly what is suggested by the large-$N$ studies
\cite{Cheung}.  However, it is not yet clear how or at what
parameter values the spiral state changes into the spin-liquid.

At lower values of $U/W$ (i.e., near the Mott transition) higher
order perturbation processes cannot be neglected. For example, ring
exchange processes (which is of order $t^4/U^3$)
\cite{MotrunichPRB05} and hopping around triangular clusters (which
is of order $t^2t'/U^2$) \cite{NaxCoO} may become important. Both of
these processes favour RVB spin-liquid states. This may explain the
observation of a spin-liquid state in insulating phase of \CN
\cite{Shimizu} and \dmit \cite{Kato} (see section \ref{sect:spinL}).
Therefore, we add an RVB spin-liquid phase to our sketch of the
phase diagram intervening between the quasi--one-dimensional phase
and the spiral order. We also include the enhancement of the region
of stability of this phase near to the Mott transition.

\section{Some outstanding problems}

Before drawing our conclusions, we briefly indulge ourselves by
discussing what we see as some of the major problems in the field
that we have not discussed significantly above.

\subsection{Deducing parameters for the minimal
models from first principles electronic structure
calculations}\label{sect:cal-params}

We have argued above that the physics of the organic charge transfer
salts can be understood in terms of the Hubbard model on the
anisotropic triangular lattice. Clearly it is vitally important, for
this program, to know accurately what values of the parameters of
this model ($t$, $t'$ and $U$) correspond to a given material at a
given pressure. Understandably, given the chemical complexity and
large unit cells of the organic charge transfer salts, most
theoretical work on the band structure of these materials has used
the Huckel model \cite{reviews}. (There has also been considerable
effort expended to determine the values of $t$ and $t'$
experimentally \cite{Singleton}.) However, recent advances in
computational speed and the computational efficiency of electronic
structure codes \cite{orderN} have allowed the first density
functional theory (DFT) studies of the band structures of organic
charge transfer salts to be performed
\cite{Lee,Miyazaki,Kino,alpha}. The calculations are significantly
more accurate than the Huckel calculations and there is a great need
for systematic studies of the band structures of a range organic
charge transfer salts. In particular a detailed mapping from the
experimental parameters of `chemical' and hydrostatic pressure to
the theoretical parameters of $t$ and $t'$ would enormously benefit
the field. Further, this would greatly improve our ability to
perform quantitative tests of theories of the Hubbard model in the
organic charge transfer salts.

To complete the mapping between the experimental and theoretical
parameters we also need to know how the dimer $U$ varies with
`chemical' and hydrostatic pressure. This is a much more difficult
task. To date the most common approach has been to estimate $U$ from
the intra-dimer hopping integral. We have shown in section
\ref{sect:hlos}, that this is not only inaccurate (as it is based on
extend Huckel calculations) but also incorrect as correlation
effects, not the intra-dimer hopping integrals, determine the dimer
$U$. Therefore, more accurate calculations are required. The `bare'
Hubbard-$U$ of a dimer \emph{in vacuo} can be straightforwardly
determined form DFT as it is the second derivative of energy with
respect to charge ($U=d^2E/dq^2$). However, in a crystal the
effective Hubbard-$U$ is greatly decreased as the second derivative
of energy with respect to charge contains a large contribution due
to the polarisability of the crystal. The problem of calculating the
reduction in the Hubbard-$U$ due to the polarisability of the
molecules is simply a self-consistent problem in classical
electrodynamics. This approach has been successfully applied to
alkali doped fullerides \cite{Gunnarsson,C60-alpha}. Thus the large
size and low electron density of molecular crystals is a singular
advantage when it comes to calculating the Hubbard-$U$. Indeed this
is such an advantage that, we believe, this makes molecular crystals
an ideal class of materials in which to study first principles
approaches to strongly correlated effects. We believe that the time
is now ripe to apply this method to the organic charge transfer
salts and thus to map from `chemical' and hydrostatic pressure to
the Hubbard-$U$.

If both of these procedures were accurately carried out then we
would have a direct mapping between the Hubbard model and
experiment. This would allow very direct testing of the hypothesis
that the Hubbard model provides the correct microscopic description
of the organic charge transfer salts. It would also allow for the
experimental testing of the various approximations used to study the
Hubbard model. Therefore, this mapping would not only be extremely
important to the community interested in the organic charge transfer
salts for their own sake, but would make the organic charge transfer
salts an even more important test-bed for theories of strongly
correlated quantum many-body systems than they are already. In
figure \ref{fig:tri-lat-pd-mat} we sketch our best guess of which
parameters correspond to which materials, based on the calculations
performed thus far and comparison of the observed behaviour of these
materials with figure \ref{fig:tri-lat-pd}. Given the comments above
this is clearly a very rough procedure.

\subsection{Is there a pseudogap?}\label{sect:pseudoTheory}

\begin{figure}[t]
\centering
\includegraphics*[width=.7\textwidth]{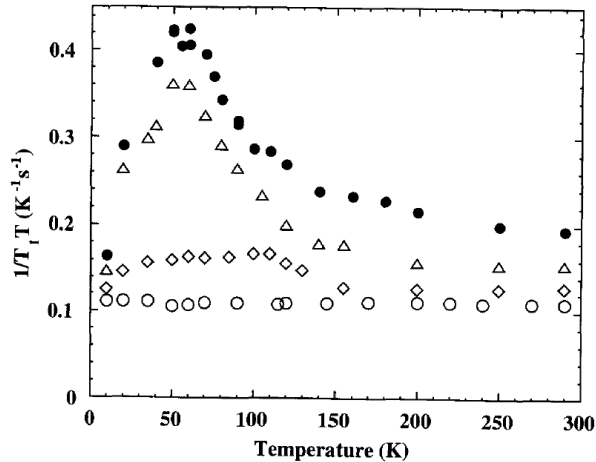}
\caption[]{Is the suppression of the nuclear spin relaxation rate,
$1/T_1$, evidence for a pseudogap? This data (from \cite{Wzietek})
shows the temperature dependence of $1/T_1T$ in \Br at various
pressures [1 bar ($\bullet$), 1.5 kbar ($\triangle$), 3 kbar
($\diamond$) and 4 kbar ($\circ$)]. Rather similar effects are seen
across a range layered organic charge transfer salts
\cite{NMRreview}. In materials that are close to the Mott transition
a peak in $1/T_1T$ is observed in at around 50~K. This is not
expected in weakly interacting systems. However, a pseudogap is
predicted by the RVB theory of superconductivity \cite{RVB-organics}
and appears to be a rather natural feature of extensions to
dynamical mean field theory which include some spatial correlations
\cite{DCA+CDMFT,arcsCDMFT}. A detailed theoretical understanding of
this data is still lacking, as are further experiments probing the
possibility that there is a pseudogap in the organic charge transfer
salts.} \label{fig:nmr}
\end{figure}

NMR experiments on the organic charge transfer salts show a large
decrease in the spin lattice relaxation rate, $1/T_1T$, the Knight
shift, $K_s$, and the Korringa ratio, $1/T_1TK_s^2$ below about 50~K
\cite{de_Soto,Mayaffre,NMRreview} (see figure \ref{fig:nmr}). In
this section we ask whether these experiments indicate the opening
of a gap-like--structure at the Fermi energy. The pseudogap phase in
the cuprates has attracted a great deal of attention \cite{Carlson},
yet there has been very little work on understanding the origin of
these experimental effects in the the organic charge transfer salts.
Clearly a good starting point to investigate this effect in the
organic charge transfer salts would be to follow a number of the
experimental and theoretical approaches that have proved fruitful in
investigating the pseudogap in the cuprates.

One of the most important questions about the pseudogap in the
cuprates has been whether, and if so how, it relates to
superconductivity. The Nernst effect is much larger in a type-II
superconductor than it is in a normal metal (because vortices carry
entropy extremely efficiently). Therefore, the discovery of a large
Nernst effect in the pseudogap regime \cite{Ong} of the cuprates is
suggestive of superconducting fluctuations playing an important role
in the pseudogap \cite{EK}. In this context studies of the
temperature dependence of the Nernst effect in materials which show
the unexpected NMR behaviour \{e.g., \Br and \NCSn\} would be
extremely interesting.

Theoretical avenues could also be productively pursued. In
particular we need to discover the correct phenomenological
description of the NMR data. The first question which needs to be
addressed is can the NMR data be explained without invoking a gap in
the density of states? For example, the maximum in $1/T_1T$ occurs
at a temperature very close to that where the crossover from the
`bad-metal' to the Fermi liquid occurs. Could the origin of the
maximum then be simply related to recovery of a Fermi-liquid
Korringa-like behaviour from the local moment (Heisenberg) physics
associated with the `bad-metal' \cite{MMP,Moriya,Eddy-private}
(charge fluctuations happen on significantly slower time-scales than
spin fluctuations in the `bad-metal')? Or, is a gap in the density
of states necessary to describe the observed phenomena? A
phenomenological description of the data will obviously provide
clues to the appropriate microscopic description. Given the success
of DMFT in describing many of properties of the organic charge
transfer salts (see section \ref{sect:DMFT}), extensions to DMFT to
include non-local effects seem an obvious avenue to explore
\cite{DCA+CDMFT}. Cellular DMFT (CDMFT) \cite{DCA+CDMFT} and the
dynamical cluster approximation (DCA) \cite{DCA+CDMFT} are two such
approaches. These approaches have already provided significant
insights into the pseudogap in the cuprates, in particular the
so-called Fermi arcs seen in angle resolved photoemission
spectroscopy (ARPES) experiments \cite{cuprate-ARPES,Norman} on the
cuprates appear to be a natural feature in CDMFT \cite{arcsCDMFT}.
Therefore, studies of the Hubbard model on an anisotropic triangular
lattice within CDMFT or the DCA would be an extremely interesting
approach.\footnote{One CDMFT study of the Hubbard model on an
anisotropic triangular lattice has appeared in the literature but
this problem was not addressed in that paper \cite{Kyung&Tremblay}.}
It is interesting to note that many of the features present in the
CDMFT description of the pseudogap are captured (in a somewhat
cruder form) by the RVB theory (discussed in section
\ref{sect:SCtheory}). In particular the RVB theory predicts a
pseudogap \cite{ZhangPRB,Zhang-organic,unpublished,RVB-organics}.
Therefore, a detailed understanding of the NMR experiments on the
organic charge transfer salts will provide a stringent test of this
theory.

\subsection{Low $\mathbf{T_c}$ materials}

Most of the attention to superconducting has been focused on
materials with high transition temperatures and the region around
the Mott transition. But, recently, the work Pratt, Blundell and
coworkers \cite{constraints,Pratt_nU,Pratt&Blundell} has shown the
the importance of materials with low critical temperatures for
understanding superconductivity in the organic charge transfer salts
(see figure \ref{fig:penetration} and section \ref{sect:Pratt} for a
discussion of this work). However, there remains very little data on
the low-$T_c$ materials. We believe that an extensive understanding
of the phenomenology of these low-$T_c$ materials is a vital
prerequisite of a more detailed understanding of the superconducting
states of \emph{all} of the organic charge transfer salts.
Therefore, there is a desperate need for more experimental and
theoretical work on these materials.

\subsection{Synthetic chemistry}

It goes, almost, without saying that condensed matter physics cannot
proceed without high quality samples. As such, synthetic chemistry
plays the central role in the study of organic charge transfer
salts, for, without it, the field could neither exist nor progress.
Historically, much of the synthetic work in this field has focused
on discovering new organic charge transfer salts. However, we would
like to stress the importance of the less obviously glamorous work
of producing higher quality samples of  materials which have already
been extensively studied. As we have seen in this review there are
many outstanding problems in the flavours of organic charge transfer
salts that have been with us for the last twenty years. Many of
these problems would be made significantly more tractable if new
kinds of data were available. For example, a tricrystalline
experiment, of the type so important in the cuprates
\cite{VanHarlingen}, could lay to rest once and for all questions
about the pairing symmetry in the organic charge transfer salts.

Perhaps the most pressing need for improvement in sample quality is
size. Neutron scattering has been extremely important for our
understanding of the cuprates \cite{cuprate-neutron}, Na$_x$CoO$_2$
\cite{NCO-neutron,NaxCoO} and many other strongly correlated
materials. This is mostly because neutrons provide direct
information about the spin degrees of freedom which are of vital
importance in strongly correlated materials such as the organic
charge transfer salts. Therefore, perhaps the single greatest
impediment to the field is the lack of high quality single crystals
for neutron studies. Taniguchi and coworkers have recently
demonstrated that large single crystals can be grown \cite{ICSM} and
one can only eagerly await the results of neutron scattering
experiments on this sample and the growth of similarly large samples
of other salts.

\subsection{Do \CN and/or \dmit have spin-liquid ground
states?}\label{sect:spinL}

\begin{figure}[t]
\centering
\includegraphics*[width=.57\textwidth]{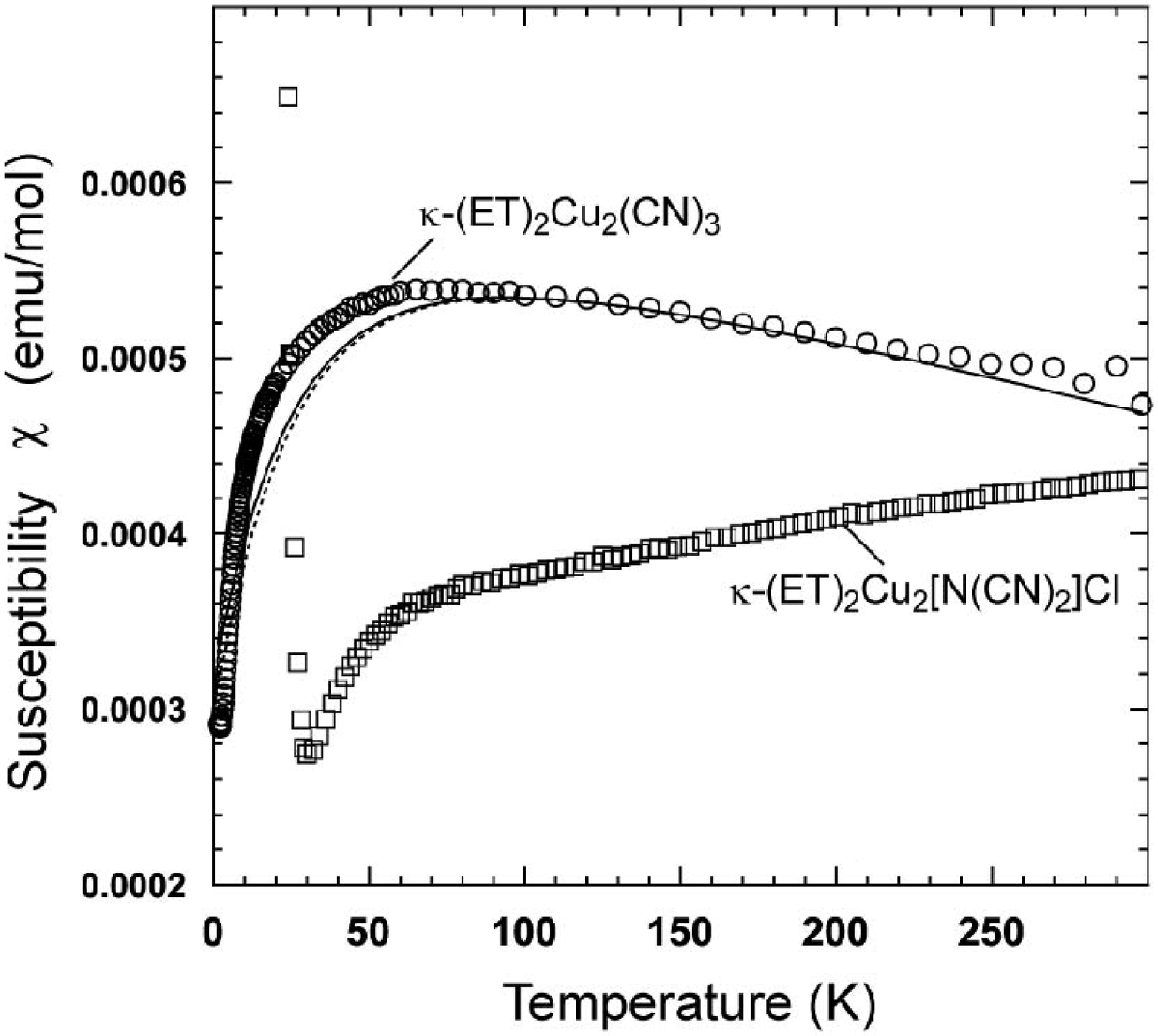}
\includegraphics*[width=.32\textwidth]{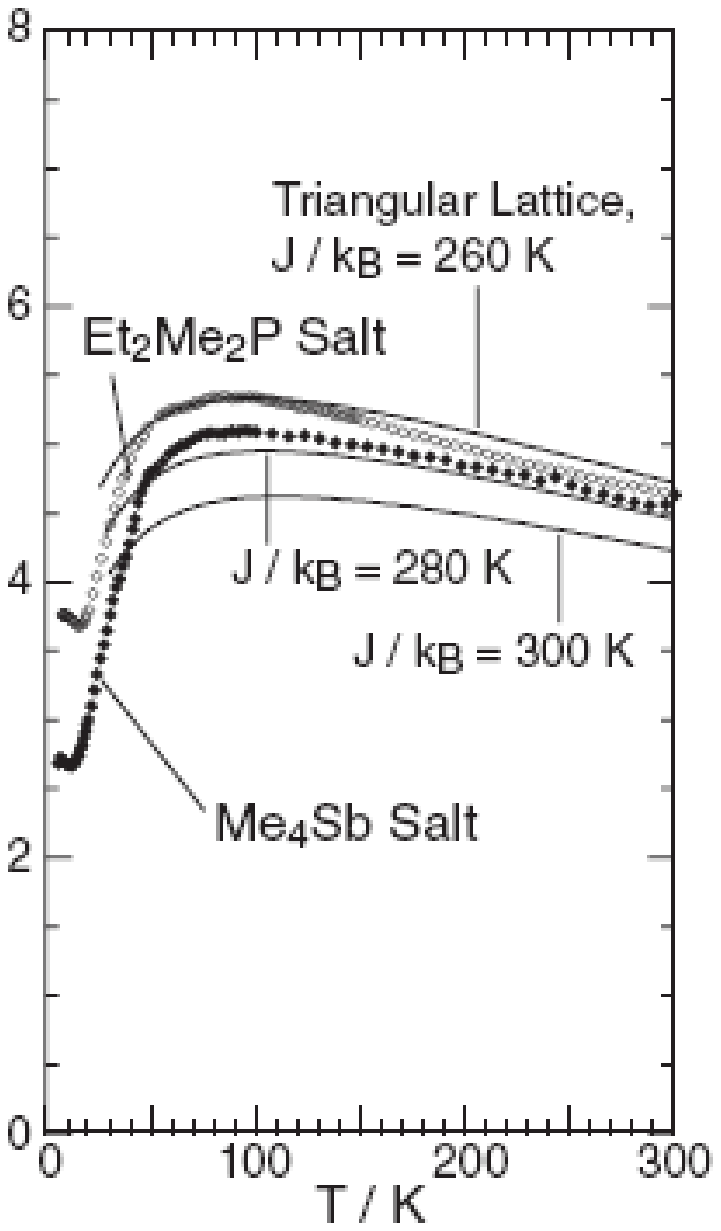}
\caption[]{Evidence for spin-liquid ground states in \CN and \dmitn.
The left panel (modified from \cite{Shimizu}) compares the
temperature dependencies of the magnetic susceptibilities, $\chi$,
of \CN and \Cln. Both measurements were performed in the ambient
pressure Mott insulating phases of the two compounds. In \Cl a rapid
upturn in the susceptibly is observed at $\sim$25~K as a result of
the N\'eel transition. Remarkably no magnetic ordering transition is
observed in \CN down to the lowest temperatures studied (32~mK).
Fits of series expansion data (lines in figure) to the measured
susceptibility \cite{Shimizu} show that the exchange interaction,
$J$, is about 250~K in both materials. This had led to the
suggestion that it is the greater frustration in \CN that leads to a
spin-liquid ground state. [In \CN $J'/J\sim1$ whereas in \Cl
$J'/J<1$.] The right panel shows similar measurements in \dmit
\cite{Kato}. This data is also well described by series expansions
for the isotropic triangular lattice (solid lines in figure with $J$
marked). Except for $X$=Et$_2$Me$_2$Sb the \dmit salts do eventually
order magnetically, but this suggests that it may be possible to use
chemistry and/or uniaxial stress tune the \dmit salts across the
antiferromagnetic/spin-liquid phase boundary. Understanding the
microscopic nature of these spin-liquid states and why they are
stabilised over the 120$^\circ$-state which appears to be the ground
state of Heisenberg model on the isotropic triangular lattice
present major challenges to theory.} \label{fig:shimizu}
\end{figure}

A recent series of experiments on \CN \cite{Shimizu-rev,Shimizu} by
Shimizu and coworkers has sparked considerable interest in the
theoretical community
\cite{unpublished,MotrunichPRB05,RVB-organics,Kyung&Tremblay,tri-series,Liu,Fjaerestad,Alicea,Raman,Nikolic,LeeLee,Singh,Parcollet,Kondo-Moriya}.
This is because Shimizu \etal have gathered strong evidence that \CN
may have a spin-liquid ground state. A spin-liquid is a state in
which there exist well formed local moments which do not
magnetically order. Importantly a spin-liquid should posses
\emph{all} of the symmetries of the crystal. In a particularly
beautiful demonstration of the spin-liquid behaviour in \CN Shimizu
\etal compared the spin susceptibility of \CN with that of \Cl (see
figure \ref{fig:shimizu}). Both \CN and \Cl have well formed local
moments with spin-$\frac12$. In both systems the Heisenberg exchange
interaction is about 250~K \cite{Ross-review,Shimizu}. However, the
frustration is significantly larger in \CN than it is in \Cl
\cite{Shimizu,tri-series,CN_Huckel}. The susceptibility of \Cl
diverges at $\sim$25~K as the material undergoes the N\'eel
transition. In contrast no magnetic transition is observed in \CN
down to 32~mK (the lowest temperature studied by Shimizu \etaln).

Rather similar behaviour (also shown in figure \ref{fig:shimizu})
has been observed by Kato \etal \cite{Kato} in a the series \dmitn,
where Pd(dmit)$_2$ is the acceptor molecule
1,3-dithiol-2-thione-4,5-dithiolate (C$_3$S$_5$), $Z$=Me$_4Y$ or
Et$_2$Me$_2Y$, $Y$=P or Sb, Me=CH$_3$ and Et=C$_2$H$_5$. So far,
this has not attracted so much attention from theorists as the \CN
results, but these materials are also excellent candidates for the
observation of spin-liquid states. The susceptibility of the \dmit
compounds is much more reminiscent of that of \CN than that of \Cln.
In particular the Et$_2$Me$_2$Sb salt shows no indications of
magnetic ordering down to 4.3~K (the lowest temperature studied thus
far, although magnetic impurities cause a Curie tail and make this
somewhat difficult to substantiate) \cite{dmit-JMC}. Further, the
range of Curie temperatures seen in the salts that do eventually
order magnetically \cite{dmit-JMC} suggests that varying the cation,
$Z$, in the \dmit series allows one to tune the frustration and thus
the proximity to the spin-liquid state.

The clear theoretical challenge is to explain why a spin-liquid
ground state is stable in these materials.\footnote{Even if the
compounds do order at some extremely low temperature the spin-liquid
state is clearly at least very energetically competitive.} Simply
appealing to the geometrical frustration inherent in the triangular
lattice is insufficient, as it is all but certain that the
120$^\circ$-state is the ground state of Heisenberg model on a
triangular lattice (see section \ref{sect:pd-tri}). A great deal of
theoretical effort has already been expended on this problem
\cite{MotrunichPRB05,RVB-organics,Kyung&Tremblay,tri-series,Liu,Fjaerestad,Alicea,Raman,Nikolic,LeeLee,Singh,Parcollet,Kondo-Moriya}.
The idea that proximity to the Mott transition allows perturbation
terms not included in the Heisenberg model to become large seems
particularly promising. However, this question is far from settled
yet and will doubtless be the basis of much debate in the future.

\section{Conclusions}

We have seen above that the predicted behaviour of the Hubbard model
on the anisotropic triangular lattice provides good qualitative
agreement with experiments on the organic charge transfer salts.
This success is not just limited to the $\kappa$ phases but the
model applies equally well to the $\beta$, $\beta'$ and $\lambda$
phases: in spite of the chemical and structural differences between
these materials the physics is essentially the same. As the exact
solution of this model is not yet known various approximation
schemes must be used to discover the true behaviour of the model.
Perhaps the most notable successes of this model are the explanation
of the metallic state and the Mott transition in terms of dynamical
mean field theory (DMFT) discussed in section \ref{sect:DMFT}.
However, recent theories of superconductivity based on the
resonating valence bond (RVB) state appear to explain many of the
features of the superconducting state. An important test of the RVB
theory will be a detailed comparison of its prediction of a
pseudogap with experiments (such as NMR or Nernst effect) on the
strongly correlated metallic state just above $T_c$. An important
challenge for this theory is to explain the small superfluid
stiffness seen far from the Mott transition, i.e., in compounds with
low critical temperatures.

There are many exciting challenges facing the field. Some are old -
like finding an experiment which decisively settles the questions
about the pairing symmetry; while others are extremely new - like
understanding the spin-liquid state in \CNn. An important challenge
is also to increase the quantitative detail of the predictions of
theory. This is required, not only because of intrinsic interest in
the organic charge transfer salts themselves, but also because the
organic charge transfer salts provide a wonderful test-bed for many
of the most important ideas in the theory of strongly correlated
materials. In particular, the Mott transition remains perhaps
\emph{the} phenomena of central importance to strongly correlated
physics and the organic charge transfer salts are one of only a
handful of systems where the Mott transition can be driven by
varying $U/W$ as Mott originally envisioned \cite{Mott} rather than
by doping. We have highlighted one possible approach to increasingly
quantitative prediction, that of parameterising minimal models,
although there are several other approaches which might be
profitable, such as LDA+DMFT \cite{Kotliar}.

We hope that this review has made it clear that the organic charge
transfer salts are a playground for quantum many-body
physics.\footnote{Although it has suggested that the organic charge
transfer salts are a ``minefield'' rather than a playground
\cite{AndersonPrivate}!} Many of the most important phenomena in
strongly correlated physics are found in the organic charge transfer
salts, for example, the Mott transition, unconventional
superconductivity, frustrated antiferromagnetism and spin-liquids.
Further, the organic charge transfer salts are exceptionally clean
systems (as is evidenced by the beautiful quantum oscillations
observed at low temperatures \cite{Singleton,QuantOsc,Wosnitza}).
But, perhaps, what makes the organic charge transfer salts most
attractive is the relative ease with which the strength of
correlations can be controlled via hydrostatic pressure or chemistry
\cite{Taniguchi-Mott,Kanoda}. Thus, the study of the organic charge
transfer salts is an important branch of strongly correlated
physics. The lessons learned from the organic charge transfer salts
are already having significant impact in other strongly correlated
systems. For example, the spin-liquid phase of Cs$_2$CuCl$_4$ has a
great deal in common with the insulating phases of \CN and \dmit
\cite{tri-series} and the `Curie-Weiss metal' seen in Na$_x$CoO$_2$
is essentially a doped analogue of the `bad metal' observed in the
organic charge transfer salts \cite{NaxCoO}. All of these phenomena
arise from the interplay of strong correlations with frustration
which can be so elegantly studied in the organic charge transfer
salts.

\section*{Acknowledgements}

We are indebted to many people for useful discussions and
collaborations without which this review would not be possible. In
particular we would like to thank James Analytis, Arzhang Ardavan,
Steve Blundell, Jim Brooks, Tony Carrington, Amalia Coldea, Martin
Dressel, John Fj\ae restad, Greg Freebairn, Hidetoshi Fukuyama,
Anthony Jacko, Michael Lang, Fumitaka Kagawa, Kazushi Kanoda, Brad
Marston, Jaime Merino, Mark Pederson, Francis Pratt, Edan Scriven,
John Singleton, Hiromi Taniguchi, Jochen Wosnitza and Eddy Yusuf. We
are grateful to Edan Scriven and Eddy Yusuf for critically reading
this manuscript. We are also indebted to the Australian Research
Council for funding our work over a number of years.

\bibliographystyle{prsty}
\bibliography{ET_review}

\begin{thebibliography}{100}

\bibitem{A&V}
{See, for example, table 1. 1 in N. Ashcroft and D. Mermin}, {\em {Solid State
  Physics}} (Thomson Learning, Singapore, 1976).

\bibitem{Landau}
L.~D. Landau, Sov. Phys. JETP {\bf 3},  920  (1957).

\bibitem{Hewson}
A.~C. Hewson, {\em {The Kondo Problem to Heavy Fermions}} (Cambridge University
  Press, Cambridge, 1997).

\bibitem{BCS}
J. Bardeen, L.~N. Cooper, and J.~R. Schrieffer, Phys. Rev. {\bf 108},  1175
  (1957).

\bibitem{LeeWenNagaosa}
P.~A. Lee, N. Nagaosa, and X.-G. Wen, Rev. Mod. Phys. {\bf 78},  17  (2006).

\bibitem{Salamon}
M.~B. Salamon and M. Jaime, Rev. Mod. Phys. {\bf 73},  583  (2001).

\bibitem{Sigrist&Ueda}
{For a review see M. Sigrist and K. Ueda}, Rev. Mod. Phys. {\bf 63},  239
  (1991).

\bibitem{Taniguchi-Mott}
H. Taniguchi, K. Kanoda, and A. Kawamoto, Phys. Rev. B {\bf 67},  014510
  (2003).

\bibitem{Sasaki}
T. Sasaki, N. Yoneyama, A. Suzuki, N. Kobayashi, Y. Ikemoto, and H. Kimura, J.
  Phys. Soc. Japan {\bf 74},  2351  (2005).

\bibitem{AndersonRVB}
P.~W. Anderson, Science {\bf 235},  1196  (1987).

\bibitem{NaxCoO}
{For a recent review see J. Merino, B. J. Powell, and R. H. McKenzie},
  cond-mat/0512696 (to appear in Phys. Rev. B).

\bibitem{alt}
{See, for example}, {A. Girlando, M. Masino, A. Brillante, R. G. DellaValle and
  E. Venuti, Phys. Rev. B \textbf{66}, 100507 (2002); G. Varelogiannis, Phys.
  Rev. Lett. \textbf{88}, 117005 (2002).}

\bibitem{heat-capacity}
H. Elsinger, J. Wosnitza, S. Wanka, J. Hagel, D. Schweitzer, and W. Strunz,
  {Phys. Rev. Lett. {\bf84}, 6098 (2000); J. M\"uller, M. Lang, R. Helfrich, F.
  Steglich, and T. Sasaki, Phys. Rev. B {\bf65}, 140509 (2002).}

\bibitem{Strack}
C. Strack, C. Akinci, V. Pashchenko, B. Wolf, E. Uhrig, W. Assmus, M. Lang, J.
  Schreuer, L. Wiehl, J.~A. Schlueter, J. Wosnitza, D. Schweitzer, J.
  M{\"u}ller, and J. Wykhoff, Phys. Rev. B {\bf 72},  54511  (2005).

\bibitem{reviews}
{See, for example}, {M. Lang and J. M\"uller, {\it Organic superconductors} in
  {\it The Physics of Superconductors - Vol. 2}, K.-H. Bennemann, J. B.
  Ketterson (Eds.), Springer-Verlag (2003); T. Ishiguro, K. Yamaji, and G.
  Saito, {\it Organic Superconductors} (Springer Verlag, Heidelberg, 1998).}

\bibitem{DFT}
A. Fortunelli and A. Painelli, {Phys. Rev. B {\bf 45}, 16088 (1997); J. Chem.
  Phys. {\bf106}, 8041 (1997); {\bf106}, 8051 (1997); S. A. French and C. R. A.
  Catlow, J. Phys. Chem. Solids {\bf65}, 39 (2004).}

\bibitem{Lee}
Y.~J. Lee, R.~M. Nieminen, P. Ordejon, and E. Canadell, Phys. Rev. B {\bf 67},
  R180505  (2003).

\bibitem{Miyazaki}
T. Miyazaki and H. Kino, Phys. Rev. B {\bf 68},  R220511  (2003).

\bibitem{Kino}
T. Miyazaki and H. Kino, Phys. Rev. B {\bf 73},  035107  (2006).

\bibitem{alpha}
{H. Kino and T. Miyazaki}, {cond-mat/0601063}.

\bibitem{Kino&Fukuyama}
H. Kino and H. Fukuyama, J. Phys. Soc. Jpn. {\bf 65},  2158  (1996).

\bibitem{quarter-filled}
J. Merino and R.~H. McKenzie, {Phys. Rev. Lett. {\bf87}, 237002 (2001); R. H.
  McKenzie, J. Merino, J. B. Marston, and O. P. Sushkov, Phys. Rev. B {\bf64},
  085109 (2001); A. Greco, J. Merino, A. Foussats, and R. H. McKenzie, Phys.
  Rev. B {\bf71}, 144502 (2005).}

\bibitem{Jaime-review}
H. Seo, J. Merino, H. Yoshioka, and M. Ogata, J. Phys. Soc. Japan {\bf 75},
  051009  (2006).

\bibitem{group}
B.~J. Powell, cond-mat/0603057.

\bibitem{Ross-review}
R.~H. McKenzie, Comments Condens. Matter Phys. {\bf 18},  309  (1998).

\bibitem{Rahal}
M. Rahal, D. Chasseau, J. Gaultier, L. Ducasse, M. Kurmoo, and P. Day, Acta
  Cryst. B {\bf 53},  159  (1997).

\bibitem{quant-chem}
{See for example}, {F. Castet, A. Fritsch, and L. Ducasse, J. Phys. I {\bf 6},
  583 (1996); L. Ducasse, A. Fritsch, and F. Castet, Synth. Met. {\bf 85}, 1627
  (1997); A. Fortunelli and A. Painelli, J. Chem. Phys. {\bf 106}, 8051 (1997);
  Y. Imamura, S. Ten-no, K. Yonemitsu, and Y. Tanimura, J. Chem. Phys.
  {\bf111}, 5986 (1999).}

\bibitem{Kanoda}
K. Kanoda, Physica C {\bf 282-287},  299  (1997).

\bibitem{Greg-honours}
G. Freebairn, ``\emph{Phonons and the Isotopically Induced Mott Transition}",
  Honours thesis, University of Queensland (2005).

\bibitem{RVB-organics}
B.~J. Powell and R.~H. McKenzie, Phys. Rev. Lett. {\bf 94},  047004  (2005).

\bibitem{Uji}
S. Uji, H. Shinagawa, T. Terashima, T. Yakabe, Y. Terai, M. Tokumoto, A.
  Kobayashi, H. Tanaka, and H. Kobayashi, Nature {\bf 410},  908  (2001).

\bibitem{Jaccarino-Peter}
V. Jaccarino and M. Peter, {Phys. Rev. Lett. {\bf9}, 290 (1962). L. Balicas, J.
  S. Brooks, K. Storr, S. Uji, M. Tokumoto, H. Tanaka, H. Kobayashi, A.
  Kobayashi, V. Barzykin, and L. P. Gor'kov, Phys. Rev. Lett. {\bf87}, 067002
  (2001).}

\bibitem{Cepas}
O. C\'{e}pas, R.~H. McKenzie, and J. Merino, Phys. Rev. B {\bf 65},  100502(R)
  (2002).

\bibitem{UjiBrooks}
S. Uji and J.~S. Brooks, J. Phys. Soc. Japan {\bf 75},  051014  (2006).

\bibitem{alpha-band}
H. Kino and T. Miyazaki, cond-mat/0601063.

\bibitem{Grant}
P.~M. Grant, Phys. Rev. B {\bf 26},  6888  (1982).

\bibitem{Dupuis}
N. Dupuis, C. Bourbonnais, and J.~C. Nickel, cond-mat/0510544.

\bibitem{Kotliar-phys-today}
{G. Kotliar and D. Vollhardt}, Phys. Today {\bf 57},  53  (2004).

\bibitem{DMFT}
G. Kotliar and D. Vollhardt, {Phys. Today {\bf57}, 53 (2004); A. Georges, G.
  Kotliar, W. Krauth, and M. Rozenberg, Rev. of Mod. Phys. {\bf68}, 125
  (1996).}

\bibitem{Wosnitza}
{For a comprehensive review see J. Wosnitza}, {\em {Fermi Surfaces of
  Low-Dimensional Organic Metals and Superconductors: Springer Tracts in Modern
  Physics {\bf134}}} (Springer Verlag, Berlin, 1996).

\bibitem{QuantOsc}
{For recent reviews see J. Singleton}, {Rep. Prog. Phys. {\bf63} 1111 (2000);
  M. Kartsovnik, Chem. Rev. {\bf104}, 5737 (2004).}

\bibitem{Bulaevskii}
L.~N. Bulaevskii, Adv. Phys. {\bf 37},  443  (1988).

\bibitem{Analytis}
J.~G. Analytis, A. Ardavan, S.~J. Blundell, R.~L. Owen, E.~F. Garman, C.
  Jeynes, and B.~J. Powell, Phys. Rev. Lett. {\bf 96},  177002  (2006).

\bibitem{Abrikosov}
{See, for example, A. A. Abrikosov}, {\em {Introduction to the theory of normal
  metals}} (Academic Press, New York, 1972).

\bibitem{Baber}
W. Baber, Proc. Roy. Soc. A {\bf 158},  383  (1937).

\bibitem{MottMin}
{N. F. Mott}, {\em {Metal insulator transitions}} (Taylor and Francis, London,
  1990).

\bibitem{Gunnarsson}
{O. Gunnarsson}, {\em {Alkali-doped fullerides: Narrow-band solids with unusual
  properties}} (World Scientific, Singapore, 2004).

\bibitem{SRO4}
A.~W. Tyler, A.~P. Mackenzie, S. Nishizaki, and Y. Maeno, Phys. Rev. B {\bf
  58},  R10107  (1998).

\bibitem{SRO3}
{L. Klein}, {Phys. Rev. Lett. {\bf77}, 2774 (1996); P. B. Allen, H. Berger, O.
  Chauvet, L. Forro, T. Jarlborg, A. Junod, B. Revaz, and G. Santi, Phys. Rev.
  B {\bf53}, 4393 (1996).}

\bibitem{VO2}
P.~B. Allen, R.~M. Wentzcovitch, W.~W. Schulz, and P.~C. Canfield, Phys. Rev. B
  {\bf 48},  4359  (1993).

\bibitem{optical}
K. Kornelsen, J.~E. Eldridge, C.~C. Homes, H.~H. Wang, and J.~M. Williams,
  {Solid State Commun. {\bf72}, 475 (1989); J. E. Eldridge, K. Kornelsen, H. H.
  Wang, J. M. Williams, A. V. S. Crouch, and D. M. Watkins, Solid State Commun.
  {\bf79}, 583 (1991); M. Tamura, H. Tajima, K. Yakushi, H. Kuroda, A.
  Kobayashi, R. Kato, and H. Kobayashi, J. Phys. Soc. Jpn. {\bf60}, 3861
  (1991); M. Dressel, J. E. Eldridge, H. H. Wang, U. Geiser, and J. M.
  Williams, Synth. Met. {\bf52}, 201 (1992); C. S. Jacobsen, J. M. Williams,
  and H. H. Wang, Solid State Commun. {\bf54}, 937 (1985); C. S. Jacobsen, D.
  B. Tanner, J. M. Williams, U. Geiser, and H. H. Wang, Phys. Rev. B {\bf35},
  9605 (1987); J. Dong, J. L. Musfeldt, J. A. Schlueter, J. M. Williams, P. G.
  Nixon, R. W. Winter, and G. L. Gard, Phys. Rev. B {\bf60}, 4342 (1999); A.
  Schwartz, M. Dressel, G. Grüuner, V. Vescoli, L. Degiorgi, and T. Giamarchi,
  Phys. Rev. B {\bf58}, 1261 (1998).}

\bibitem{Limelette}
P. Limelette, P. Wzietek, S. Florens, A. Georges, T.~A. Costi, C. Pasquier, D.
  J\'erome, C. M\'ezi\`ere, and P. Batail, Phys. Rev. Lett. {\bf 91},  016401
  (2003).

\bibitem{Yu}
R.~C. Yu, J.~M. Williams, H.~H. Wang, J.~E. Thompson, A.~M. Kini, K.~D.
  Carlson, J. Ren, M.-H. Whangbo, and P.~M. Chaikin, Phys. Rev. B {\bf 44},
  6932  (1991).

\bibitem{JaimeDMFT}
J. Merino and R.~H. McKenzie, Phys. Rev. B {\bf 61},  7996  (2000).

\bibitem{Ramirez}
{A. P. Ramirez}, {Annu. Rev. Mater. Sci. {\bf 24}, 453 (1994); P. Schiffer and
  I. Daruka, Phys. Rev. B {\bf56}, 13712 (1997).}

\bibitem{tri-series}
W. Zheng, R.~R.~P. Singh, R.~H. McKenzie, and R. Coldea, Phys. Rev. B {\bf 71},
   134422  (2005).

\bibitem{Cv}
B. Andraka, J.~S. Kim, G.~R. Stewart, K.~D. Carlson, H.~H. Wang, and J.~M.
  Williams, {Phys. Rev. B {\bf40}, 11345 (1989); B. Andraka, C. S. Jee, J. S.
  Kim, G. R. Stewart, K. D. Carlson, H. H. Wang, A. V. S. Crouch, A. M. Kini,
  and J. M. Williams, Solid State Comm. {\bf79} 57 (1991).}

\bibitem{Huckel}
R.~C. Haddon, A.~P. Ramirez, and S.~H. Glarum, Adv. Mater. {\bf 6},  316
  (1994).

\bibitem{Frikach}
K. Frikach, M. Poirier, M. Castonguay, and K.~D. Truong, Phys. Rev. B {\bf 61},
   R6491  (2000).

\bibitem{Fournier}
D. Fournier, M. Poirier, M. Castonguay, and K.~D. Truong, Phys. Rev. Lett. {\bf
  90},  127002  (2003).

\bibitem{Jaime-HH-DMFT}
J. Merino and R.~H. McKenzie, Phys. Rev. B {\bf 24},  16442  (2000).

\bibitem{Georges-HH-DMFT}
S.~R. Hassan, A. Georges, and H.~R. Krishnamurthy, Phys. Rev. Lett. {\bf 94},
  036402  (2005).

\bibitem{Ross-Wilson}
R.~H. McKenzie, cond-mat/9905044.

\bibitem{Hussey}
N.~E. Hussey, J. Phys. Soc. Japan {\bf 74},  1107  (2005).

\bibitem{Jacko}
A. Jacko, private communication (2006).

\bibitem{T2phonons}
{M. Weger}, {J. Low Temp. Phys. {\bf95}, 131 (1994); M. Weger and D.
  Schweitzer, Synth. Met. {\bf70}, 889 (1995); J. Hagel, J. Wosnitza, C.
  Pfleiderer, J. A. Schlueter, J. Mohtasham, and G. L. Gard, Phys. Rev. B
  {\bf68}, 104504 (2003).}

\bibitem{first-order}
R. Bulla, T.~A. Costi, and D. Vollhardt, Phys. Rev. B {\bf 64},  045103
  (2001).

\bibitem{LimeletteVO}
P. Limelette, A. Georges, D. Jérome, P. Wzietek, P. Metcalf, and J.~M. Honig,
  Science {\bf 302},  89  (2003).

\bibitem{Goldenfeld}
{N. Goldenfeld}, {\em {Lectures on Phase Transitions and the Renormalization
  Group}} (Westview, Boulder, 1992).

\bibitem{Kagawa}
F. Kagawa, K. Miyagawa, and K. Kanoda, Nature {\bf 534},  89  (2005).

\bibitem{Imada}
M. Imada, {Phys. Rev. B {\bf72}, 075113 (2005); J. Phys. Soc. Jpn. {\bf74}, 859
  (2005).}

\bibitem{Misawa}
T. Misawa, Y. Yamaji, and M. Imada, {cond-mat/0604387.}

\bibitem{graphite}
T.~E. Weller, M. Ellerby, S.~S. Saxena, R.~P. Smith, and N.~T. Skipper, Nature
  Phys. {\bf 1},  39  (2005).

\bibitem{Mico}
A.~P. Micolich, E. Tavener, B.~J. Powell, A.~R. Hamilton, M. Curry, R. Geidd,
  and P. Meredith, cond-mat/0509278.

\bibitem{Eliashberg}
{G. M. Eliashberg}, {JETP {\bf38}, 966 (1960); JETP {\bf39}, 1437 (1960)}.

\bibitem{SCDFT}
M. L{\"u}ders, M.~A.~L. Marques, N.~N. Lathiotakis, A. Floris, G. Profeta, L.
  Fast, A. Continenza, S. Massidda, and E.~K.~U. Gross, {Phys. Rev. B {\bf72},
  024545 (2005); M. A. L. Marques, M. L{\"u}ders, N. N. Lathiotakis, G.
  Profeta, A. Floris, L. Fast, A. Continenza, E. K. U. Gross, and S. Massidda,
  Phys. Rev. B {\bf72}, 024546 (2005).}

\bibitem{Ketterson&Song}
{See, for example, J.B. Ketterson and S.N. Song}, {\em {Superconductivity}}
  (Cambridge University Press, Cambridge, 1999).

\bibitem{James_adv_phys}
J.~F. Annett, Adv. Phys. {\bf 39},  83  (1990).

\bibitem{Tinkham-group}
{See, for example, M. Tinkham}, {\em {Group Theory and Quantums Mechanics}}
  (McGraw-Hill, New York, 1964).

\bibitem{Morse}
R.~W. Morse, T. Olsen, and J.~D. Gavenda, Phys. Rev. Lett. {\bf 3},  15
  (1959).

\bibitem{Leggett}
A.~J. Leggett, Rev. Mod. Phys. {\bf 37},  331  (1975)).

\bibitem{Vollhardt}
D. Vollhardt and P. W{\"o}lfle, {\em {The Superfluid Phases of Helium 3}}
  (Taylor and Francis, London, 1990).

\bibitem{de_Soto}
S.~M. deSoto, C.~P. Slichter, A.~M. Kini, H.~H. Wang, U. Geiser, and J.~M.
  Williams, Phys. Rev. B {\bf 52},  10364  (1995).

\bibitem{Mayaffre}
H. Mayaffre, P. Wzietek, D. J\'erome, C. Lenoir, and P. Batail, Phys. Rev.
  Lett. {\bf 75},  4122  (1995).

\bibitem{NMRreview}
K. Miyagawa, K. Kanoda, and A. Kawamoto, Chem. Rev. {\bf 104},  5635  (2004).

\bibitem{Balian&Werthamer}
R. Balian and N.~R. Werthamer, Phys. Rev. {\bf 131},  1553  (1963).

\bibitem{Zuo}
F. Zuo, J.~S. Brooks, R.~H. McKenzie, J.~A. Schlueter, and J.~M. Williams,
  Phys. Rev. B {\bf 61},  750  (2000).

\bibitem{Clogston}
A.~M. Clogston, Phys. Rev. Lett. {\bf 9},  266  (1962).

\bibitem{Chandrasekhar}
B.~S. Chandrasekhar, Appl. Phys. Lett. {\bf 1},  7  (1962).

\bibitem{Ben3}
{B. J. Powell, J. F. Annett, and B. L. Gy{\"o}rffy}, {J. Phys. A {\bf36}, 9289
  (2003); J. Phys.: Condens. Matter {\bf15}, L235 (2003).}

\bibitem{Murata87}
K. Murata, M. Tokumoto, H. Anzai, K. Kajimura, and T. Ishiguro, J. Appl. Phys
  {\bf 26},  1367  (1987).

\bibitem{Nam}
M.-S. Nam, J.~A. Symington, J. Singleton, S.~J. Blundell, A. Ardavan, J.~A.
  A.~J. Perenboom, M. Kurmoo, and P. Day, J. Phys.: Condens. Matter {\bf 11},
  477  (1999).

\bibitem{Izawa_ET}
K. Izawa, H. Yamaguchi, T. Sasaki, and Y. Matsuda, Phys. Rev. Lett. {\bf 88},
  27002  (2002).

\bibitem{q1d-phenom}
B.~J. Powell, cond-mat/0606188.

\bibitem{Carrington}
A. Carrington, I.~J. Bonalde, R. Prozorov, R.~W. Giannetta, A.~M. Kini, J.
  Schlueter, H.~H. Wang, U. Geiser, and J.~M. Williams, Phys. Rev. Lett. {\bf
  83},  4172  (1999).

\bibitem{disorder}
B.~J. Powell and R.~H. McKenzie, Phys. Rev. B {\bf 69},  024519  (2004).

\bibitem{AndersonTheorem}
P.~W. Anderson, J. Phys. Chem. Solids {\bf 11},  26  (1959).

\bibitem{VanHarlingen}
{D. J. van Harlingen}, Rev. Mod. Phys. {\bf 67},  515  (1995).

\bibitem{Davis}
{K. M. Lang, V. Madhavan, J. E. Hoffman, E. W. Hudson, H. Elsakl, S. Uchida,
  and J. C. Davis}, {Nature {\bf415}, 412 (2002); E. W. Hudson, K. M. Lang, V.
  Madhavan, S. H. Pan, H. Elsakl, S. Uchida, and J. C. Davis, Nature {\bf411},
  920 (2001); S. H. Pan, E. W. Hudson, K. M. Lang, H. Elsakl, S. Uchida, and J.
  C. Davis, Nature {\bf403}, 746 (2000)}.

\bibitem{Mineev&Samokhin}
{See, for example, V.P. Mineev and K.V. Samokhin}, {\em {Introduction to
  Unconventional Superconductivity}} (Gordon and Breach, Amsterdam, 1998).

\bibitem{Larkin}
A. Larkin, JETP Lett. {\bf 2},  130  (1965).

\bibitem{AG}
A.~A. Abrikosov and L.~P. Gorkov, Sov. Phys. JETP {\bf 12},  1243  (1961).

\bibitem{Abrikosov-nobel}
A.~A. Abrikosov, Rev. Mod. Phys. {\bf 76},  975  (2004).

\bibitem{Annett}
{See, for example, J. F. Annett}, {\em {Superconductivity, Superfluids and
  Condensates}} (Oxford University Press, Oxford, 2004).

\bibitem{PrattICSM04}
{See, for example, F.L. Pratt, S.J. Blundell, T. Lancaster, M.L. Brooks, S.L.
  Lee, N. Toyota and T. Sasaki}, {Synthetic Metals {\bf152}, 417 (2005) and
  references therein.}

\bibitem{Joynt&Tallefer}
{For a recent review see R. Joynt and L. Taillefer}, Rev. Mod. Phys. {\bf 74},
  235  (2002).

\bibitem{Meano&Mackenzie}
{For a recent review see A.P. Mackenzie and Y. Maeno}, Rev. Mod. Phys. {\bf
  75},  657  (2003).

\bibitem{Volovik}
{G.E. Volovik}, JETP Lett. {\bf 58},  469  (1993).

\bibitem{Kubert}
{C. K{\"u}bert and P.J. Hirschfeld}, Solid State Commun. {\bf 105},  459
  (1998).

\bibitem{Vekhter}
I. Vekter, P. Hirschfeld, J. Carbotte, and E. Nicol, {Phys. Rev. B {\bf59},
  R9023 (1999); I. Vekter, P.J. Hirschfeld, and E.J. Nicol, Phys. Rev. B
  {\bf64}, 60513 (2001).}

\bibitem{Bulaevskii99}
L.~N. Bulaevskii, M.~J. Graf, and M.~P. Maley, Phys. Rev. Lett. {\bf 83},  388
  (1999).

\bibitem{Deguchi}
K. Deguchi, Z. Mao, H. Yaguchi, and Y. Maeno, Phys. Rev. Lett. {\bf 92},  47002
   (2004).

\bibitem{Won&Maki}
{H. Won and K. Maki}, Europhys. Lett. {\bf 30},  421  (1995).

\bibitem{Simon}
{S.H. Simon and P.A. Lee}, Phys. Rev. Lett. {\bf 78},  1548  (1997).

\bibitem{YBCO_Cv}
K.~A. Moler, D.~J. Baar, J.~S. Urbach, R. Liang, W.~N. Hardy, and A.
  Kapitulnik, {Phys. Rev. Lett. {\bf73}, 2744 (1994); R.A. Fisher, J.E. Gordon,
  S.F. Reklis, D.A. Wright, J.P. Emerson, B.F. Woodfield, E.M. McCarron III,
  and N.E. Phillips, Physica C {\bf252}, 237 (1995); B. Revaz, J.-Y. Genoud, A.
  Junod, K. Neumaier, A. Erb, and E. Walker, Phys. Rev. Lett. {\bf80}, 3364
  (1998); D. A. Wright, J. P. Emerson, B. F. Woodfield, J. E. Gordon, R. A.
  Fisher, and N. E. Phillips, Phys. Rev. Lett. {\bf82}, 1550 (1999).}

\bibitem{Fisher}
R. Fisher, N. Phillips, A. Schilling, B. Buffeteau, R. Calemczuk, T.
  Hargreaves, C. Marcenat, K. Dennis, R. McCallum, and A. O'Connor, Phys. Rev.
  B {\bf 61},  1473  (2000).

\bibitem{Maki}
{K. Maki}, Phys. Rev. {\bf 158},  397  (1967).

\bibitem{Lowell}
{J. Lowell and J.B. Sousa}, J. Low Temp. Phys. {\bf 3},  65  (1970).

\bibitem{Aubin}
H. Aubin, K. Behnia, M. Ribault, R. Gagnon, and L. Taillefer, Phys. Rev. Lett.
  {\bf 78},  2624  (1997).

\bibitem{Moreno}
J. Moreno and P. Coleman, Phys. Rev. B {\bf 53},  2995  (1996).

\bibitem{Suderow}
H. Suderow, H. Aubin, K. Behnia, and A. Huxley, Phys. Lett. A {\bf 234},  64
  (1997).

\bibitem{Izawa_SRO}
M.~A. Tanatar, S. Nagai, Z.~Q. Mao, Y. Maeno, and T. Ishiguro, {Phys. Rev. B
  {\bf 63}, 064505 (2001); M. A. Tanatar, M. Suzuki, S. Nagai, Z. Q. Mao, Y.
  Maeno, and T. Ishiguro, Phys. Rev. Lett. {\bf86}, 2649 (2001); K. Izawa, H.
  Takahashi, H. Yamaguchi, Y. Matsuda, M. Suzuki, T. Sasaki, T. Fukase, Y.
  Yoshida, R. Settai, and Y. Onuki, Phys. Rev. Lett. {\bf86}, 2653 (2001).}

\bibitem{Dahm}
T. Dahm, H. Won, and K. Maki, {cond-mat/0006301}.

\bibitem{Tantar_BETS}
M.~A. Tanatar, T. Ishiguro, H. Tanaka, and H. Kobayashi, {Phys. Rev. B {\bf66},
  134503 (2002); M.A. Tanatar, M. Suzuki, T. Ishiguro, H. Tanaka, H. Fujiwara,
  H. Kobayashi, T. Toito, and J. Yamada, Synth. Met. {\bf137}, 1291 (2003);
  M.A. Tanatar, M. Suzuki, T. Ishiguro, H. Fujiwara, and H. Kobayashi, Physica
  C {\bf388-389}, 613 (2003); M.A. Tanatar, T. Ishiguro, H. Tanaka, and H.
  Kobayashi, Synth. Met. {\bf133-134}, 215 (2003).}

\bibitem{Tinkham}
M. Tinkham, {\em {Introduction to Superconductivity}} (McGraw-Hill, New York,
  1975).

\bibitem{Ellman}
B. Ellman, L. Taillefer, and M. Poirier, Phys. Rev. Lett. {\bf 54},  9043
  (1996).

\bibitem{Shivaram}
B. Shivaram, Y. Jeong, T. Rosenbaum, and D. Hinks, Phys. Rev. Lett. {\bf 56},
  1078  (1986).

\bibitem{Talifer}
C. Lupien, W. MacFarlane, C. Proust, L. Taillefer, Z. Mao, and Y. Meano, {Phys.
  Rev. Lett {\bf86}, 5986 (2001); H. Matsui, Y. Yoshida, A. Mukai, R. Settai,
  Y. Onuki, H. Takei, N. Kumura, H. Aoki, and N. Toyota, J. Phys. Soc. Jpn.
  {\bf69}, 3769 (2000); H. Matsui, Y. Yoshida, A. Mukai, R. Settai, Y. Onuki,
  H. Takei, N. Kumura, H. Aoki, and N. Toyota, Phys. Rev. B {\bf63}, R60505
  (2001).}

\bibitem{ultraNCS}
M. Yoshizawa, T. Kumada, and N. Yoshimoto, J. Low Temp. Phys. {\bf 105},  1745
  (1996).

\bibitem{Simizu}
T. Simizu, N. Yoshimoto, M. Nakamura, and M. Yoshizawa, Physica B {\bf
  281\&282},  896  (2000).

\bibitem{ultraBr}
K. Frikach, P. Fertey, M. Poirier, and K. Troung, Synth. Met. {\bf 103},  2081
  (1999).

\bibitem{Uemura}
Y.~J. Uemura, G.~M. Luke, B.~J. Sternlieb, J.~H. Brewer, J.~F. Carolan, W.~N.
  Hardy, R. Kadono, J.~R. Kempton, R.~F. Kiefl, S.~R. Kreitzman, P. Mulhern,
  T.~M. Riseman, D.~L. Williams, B.~X. Yang, S. Uchida, H. Takagi, J.
  Gopalakrishnan, A.~W. Sleight, M.~A. Subramanian, C.~L. Chien, M.~Z. Cieplak,
  G. Xiao, V.~Y. Lee, B.~W. Statt, C.~E. Stronach, W.~J. Kossler, and X.~H. Yu,
  Phys. Rev. Lett. {\bf 62},  2317  (1989).

\bibitem{EK}
V.~J. Emery and S.~A. Kivelson, Nature {\bf 374},  434  (1995).

\bibitem{Zhang}
F.~C. Zhang, C. Gross, T.~M. Rice, and H. Shiba, Supercond. Sci. Technol. {\bf
  1},  36  (1988).

\bibitem{Balazs}
Z. Szotek, B.~L. Gy{\"o}rffy, and W.~M. Temmerman, Phys. Rev. B {\bf 62},  3997
   (2000).

\bibitem{Mook}
H. Mook, Phys. Rev. Lett. {\bf 32},  1167  (1974).

\bibitem{Griffin}
{For a recent review see, for example, $\S$4 of A. Griffin}, {\em {Excitations
  in a Bose-Condensed Liquid}} (Cambridge University Press, Cambridge, 1993).

\bibitem{Pratt_nU}
F.~L. Pratt, S.~J. Blundell, I.~M. Marshall, T. Lancaster, S.~L. Lee, A. Drew,
  U. Divakar, H. Matsui, and N. Toyota, Polyhedron {\bf 22},  2307  (2003).

\bibitem{Lang_penetration2}
M. Lang, N. Toyota, T. Sasaki, and H. Sato, Phys. Rev. B {\bf 46},  5822
  (1992).

\bibitem{Larkin_penetration_depth}
M.~I. Larkin, A. Kinkhabwala, Y.~J. Uemura, Y. Sushko, and G. Saito, Phys. Rev.
  B {\bf 64},  144514  (2001).

\bibitem{constraints}
B.~J. Powell and R.~H. McKenzie, J. Phys.: Condens. Matter {\bf 16},  L367
  (2004).

\bibitem{Pratt&Blundell}
F.~L. Pratt and S.~J. Blundell, Phys. Rev. Lett. {\bf 94},  097006  (2005).

\bibitem{Kuroki}
{K. Kuroki}, j. Phys. Soc. Japan, {\bf75}, 051013 (2006).

\bibitem{unpublished}
B.~J. Powell and R.~H. McKenzie, cond-mat/0607079.

\bibitem{Liu}
J. Liu, J. Schmalian, and N. Trivedi, Phys. Rev. Lett. {\bf 94},  127003
  (2005).

\bibitem{Zhang-organic}
J.~Y. Gan, Y. Chen, Z.~B. Su, and F.~C. Zhang, Phys. Rev. Lett. {\bf 94},
  067005  (2005).

\bibitem{Younoki}
{S. Yunoki and S. Sorella}, {cond-mat/0602180.}

\bibitem{Sahebsara}
{P. Sahebsara and D. S{\'e}n{\'e}chal}, cond-mat/0604057.

\bibitem{Kyung&Tremblay}
{B. Kyung and A.-M. S. Tremblay}, cond-mat/0604377.

\bibitem{Laughlin}
{R. Laughlin}, cond-mat/0209269.

\bibitem{ZhangPRB}
J. Gan, F.~C. Zhang, and Z.~B. Su, Phys. Rev. B {\bf 71},  014508  (2005).

\bibitem{Brinkman-Rice}
W.~F. Brinkman and T.~M. Rice, Phys. Rev. B {\bf 2},  4302  (1970).

\bibitem{Lieb&Wu}
E.~H. Lieb and F.~Y. Wu, Phys. Rev. Lett. {\bf 20},  1445  (1968).

\bibitem{Picket-DMFT}
K. Aryanpour, W.~E. Pickett, and R.~T. Scalettar, cond-mat/0604609.

\bibitem{exact}
M. Capone, L. Capriotti, F. Becca, and S. Caprara, Phys. Rev. B {\bf 63},
  085104  (2001).

\bibitem{series-expanisions}
Z. Weihong, R.~H. McKenzie, and R.~R.~P. Singh, Phys. Rev. B {\bf 59},  14367
  (1999).

\bibitem{Jaime-spin-wave}
J. Merino, R.~H. McKenzie, J.~B. Marston, and C.~H. Chung, J. Phys.: Condens.
  Matter {\bf 11},  2965  (1999).

\bibitem{Cheung}
C.~H. Chung, J.~B. Marston, and R.~H. McKenzie, J. Phys.: Condens. Matter {\bf
  13},  5159  (2001).

\bibitem{chubukov}
A. Chubukov, S. Sachdev, and T. Senthil, J. Phys.: Condens. Matter {\bf 6},
  8891  (1994).

\bibitem{120}
B. Bernu, P. Lecheminant, C. Lhuillier, and L. Pierre, {Phys. Rev. B {\bf 50},
  10048 (1994); L. Capriotti, A. E. Trumper, and S. Sorella Phys. Rev. Lett.
  {\bf 82}, 3899 (1999); D. Huse and V. Elser, Phys. Rev. Lett. {\bf 60}, 2531
  (1988); D. J. J. Farnell, R. F. Bishop, and K. A. Gernoth, Phys. Rev. B
  \textbf{63}, 220402 (2001); R. R. P. Singh and D. A. Huse, Phys. Rev. Lett.
  {\bf 68}, 1766 (1992).}

\bibitem{MotrunichPRB05}
O.~I. Motrunich, Phys. Rev. B {\bf 72},  045105  (2005).

\bibitem{Ross-science}
R.~H. McKenzie, Science {\bf 278},  820  (1997).

\bibitem{CCCmeasureJ}
R. Coldea, D.~A. Tennant, K. Habicht, P. Smeibidl, C. Wolters, and Z.
  Tylczynski, Phys. Rev. Lett. {\bf 88},  137203  (2002).

\bibitem{CCCspin-liquid}
R. Coldea, D.~A. Tennant, A.~M. Tsvelik, and Z. Tylczynski, Phys. Rev. Lett.
  {\bf 86},  1335  (2001).

\bibitem{Shimizu}
Y. Shimizu, K. Miyagawa, K. Kanoda, M. Maesato, and G. Saito, Phys. Rev. Lett.
  {\bf 91},  107001  (2003).

\bibitem{Huckel-beta}
H. Kobayashi, R. Kato, and A. Kobayashi, synth. Met. {\bf19}, 263 (1987); T.
  Mori and H. Sasaki, Solid State Commun. {\bf62}, 525 (1987).

\bibitem{Marston}
S.-W. Tsai and J.~B. Marston, Can. J. Phys. {\bf 79},  1463  (2001).

\bibitem{Tsvelik}
{See, for example, A. M. Tsvelik}, {\em {Quantum Field Theory in Condensed
  Matter Physics}} (Cambridge Univ. Press, Cambridge, 2003).

\bibitem{Estler-Balents}
{M. Bocquet, F. H. L. Essler, A. M. Tsvelik, and A. O. Gogolin}, {Phys. Rev. B
  {\bf64}, 094425 (2001); O. A. Starykh and L. Balents Phys. Rev. Lett.
  {\bf93}, 127202 (2004).}

\bibitem{Kato}
M. Tamura and R. Kato, J. Phys.: Condens. Matter {\bf 14},  L729  (2002).

\bibitem{Singleton}
{N. Harrison, E. Rzepniewski, J. Singleton, P. J. Gee, M. M. Honold, P. Day,
  and M. Kurmoo}, j. Phys.: Condens. Matter {\bf11}, 7227 (1999); J. Caulfield,
  W. Lubczynski, F. L. Pratt, J. Singleton, D. Y. K. Ko, W. Hayes, M. Kurmoo
  and P. Day, J. Phys.: Condens. Matter {\bf6} 2911 (1994).

\bibitem{orderN}
S. Goedecker, Rev. Mod. Phys. {\bf 71},  1085  (1999).

\bibitem{C60-alpha}
{M. R. Pederson and A. A. Quong}, {Phys. Rev. B {\bf46}, 13584 (1992); V. P.
  Antropov, O. Gunnarsson, and F. Reuse, Phys. Rev. B {\bf46}, 13647 (1992).}

\bibitem{Carlson}
E.~W. Carlson, V.~J. Emery, S.~A. Kivelson, and D. Orgad,  in {\em {The Physics
  of Superconductors - Vol. 2}}, edited by K.-H. Bennemann and J.~B. Ketterson
  (Springer Verlag, Berlin, 2003).

\bibitem{Wzietek}
P. Wzietek, H. Mayaffre, D. J\'erome, and S. Brazovskii, J. Phys. I {\bf 6},
  2011  (1996).

\bibitem{DCA+CDMFT}
T. Maier, M. Jarrell, T. Pruschke, and M.~H. Hettler, Rev. Mod. Phys. {\bf 77},
   1027  (2005).

\bibitem{arcsCDMFT}
{E.C. Carter and A. J. Scholfield}, {Phys. Rev. B {\bf70}, 045107 (2004); S. S.
  Kancharla, M. Civelli, M. Capone, B. Kyung, D. S´en´echal, G. Kotliar,
  A.-M.S. Tremblay, cond-mat/0508205.}

\bibitem{Ong}
{For a recent review see Y. Wang, L. Li, and N. P. Ong}, cond-mat/0510470.

\bibitem{MMP}
A. Millis, H. Monien, and D. Pines, Phys. Rev. B {\bf 42},  1671  (1990).

\bibitem{Moriya}
T. Moriya and K. Ueda, Adv. Phys. {\bf 49},  555  (2000).

\bibitem{Eddy-private}
E. Yusuf, private communication (2006).

\bibitem{cuprate-ARPES}
J.~M. Harris, Z.-X. Shen, P.~J. White, D.~S. Marshall, M.~C. Schabel, J.~N.
  Eckstein, and I. Bozovic, {Phys. Rev. B {\bf54}, R15665 (1996). H. Ding, T.
  Yokoya, J. C. Campuzano, T. Takahashi, M. Randeria, M. R. Norman, and T. M.
  K. H. J. Giapintzakis, Nature {\bf382}, 51 (1996).}

\bibitem{Norman}
{A. Kanigel, M. R. Norman, M. Randeria, U. Chatterjee, S. Suoma, A. Kaminski,
  H. M. Fretwell, S. Rosenkranz, M. Shi, T. Sato, T. Takahashi, Z. Z. Li, H.
  Raffy, K. Kadowaki, D. Hinks, L. Ozyuzer, and J. C. Campuzano},
  {cond-mat/0605499}.

\bibitem{cuprate-neutron}
D. Basov, H.~A. Mook, B. Dabrowski, and T. Timusk, {Phys. Rev. B {\bf52},
  R13141 (1995); M. Arai, T. Nishijima, Y. Endoh, T. Egami, S. Tajima, K.
  Tomimoto, Y. Shiohara, M. Takahashi, A. Garrett, and S. M. Bennington, Phys.
  Rev. Lett. {\bf 83}, 608 (1999).}

\bibitem{NCO-neutron}
A. Boothroyd, R. Coldea, D.~A. Tennant, D. Prabhakaran, L. Helme, and C.~D.
  Frost, {Phys. Rev. Lett. {\bf 92}, 197201 (2004); S. P. Bayrakci, I.
  Mirebeau, P. Bourges, Y. Sidis, M. Enderle, J. Mesot, D. P. Chen, C. T. Lin,
  and B. Keimer, Phys. Rev. Lett. {\bf94}, 157205 (2005); L. M. Helme A. T.
  Boothroyd, R. Coldea, D. Prabhakaran, D. A. Tennant, A. Hiess, and J. Kulda,
  Phys. Rev. Lett. {\bf94}, 157206 (2005).}

\bibitem{ICSM}
{H. Taniguchi, R. Sato, K. Satoh, A. Kawamoto, H. Okamoto, T. Kobayasi, and K.
  Mizuno}, {to appear in the proceedings of the Sixth International Symposium
  on Crystalline Organic Metals, Superconductors, and Ferromagnets (ISCOM)
  September 11-16, 2005, Key West, Florida.}

\bibitem{Shimizu-rev}
Y. Shimizu, K. Miyagawa, K. Kanoda, M. Maesato, and G. Saito, Prog. Theo. Phys.
  Sup. {\bf 159},  52  (2005).

\bibitem{Fjaerestad}
W. Zheng, J.~O. Fj{\ae}restad, R.~R.~P. Singh, R.~H. McKenzie, and R. Coldea,
  Phys. Rev. Lett. {\bf 96},  057201  (2006).

\bibitem{Alicea}
J. Alicea, O.~I. Motrunich, M. Hermele, and M.~P.~A. Fisher, Phys. Rev. B {\bf
  72},  064407  (2005).

\bibitem{Raman}
K.~S. Raman, R. Moessner, and S.~L. Sondhi, Phys. Rev. B {\bf 72},  064413
  (2005).

\bibitem{Nikolic}
P. Nikoli{\'c}, Phys. Rev. B {\bf 72},  064423  (2005).

\bibitem{LeeLee}
S.-S. Lee and P.~A. Lee, Phys. Rev. Lett. {\bf 95},  036403  (2005).

\bibitem{Singh}
A. Singh, Phys. Rev. B {\bf 71},  214406  (2005).

\bibitem{Parcollet}
O. Parcollet, G. Biroli, and G. Kotliar, Phys. Rev. Lett. {\bf 92},  226402
  (2004).

\bibitem{Kondo-Moriya}
H. Kondo and T. Moriya, J. Phys. Soc. Japan {\bf 73},  812  (2004).

\bibitem{CN_Huckel}
T. Komatsu, N. Matsukawa, T. Inoue, and G. Saito, J. Phys. Soc. Japan {\bf 65},
   1340  (1996).

\bibitem{dmit-JMC}
T. Nakamura, T. Takahashi, S. Aonuma, and R. Kato, J. Mater. Chem. {\bf 11},
  2159  (2001).

\bibitem{Mott}
N.~F. Mott, Proc. Phys. Soc., London, Sect. A {\bf 62},  416  (1949).

\bibitem{Kotliar}
C.~A. Marianetti, G. Kotliar, and G. Ceder, Nature Materials {\bf 3},  627
  (2004).

\bibitem{AndersonPrivate}
{P. W. Anderson}, {private communication (2001).}

\end{thebibliography}

\end{document}